# Breaking the Low Concentration Barrier of Single-Molecule Fluorescence Quantification to the Sub-Picomolar Range

Malavika Kayyil Veedu,[1] Jérôme Wenger[1],*

[1] Aix Marseille Univ, CNRS, Centrale Med, Institut Fresnel, AMUTech, 13013 Marseille, France

*Corresponding author: jerome.wenger@fresnel.fr

**Abstract:**

Single-molecule fluorescence techniques provide exceptional sensitivity to probe biomolecular interactions. However, their application to accurately quantify analytes at the picomolar concentrations relevant for biosensing remains challenged by a severe degradation in the signal-to-background ratio. This so-called "low concentration barrier" is a major factor hindering the broad application of single-molecule fluorescence to biosensing. Here we break into the low concentration limit while keeping intact the confocal microscope architecture and without requiring complex microfluidics or preconcentration stages. Using fluorescence lifetime correlation spectroscopy (FLCS) and adding a diaphragm to the laser excitation beam, we achieve a limit of quantitation (LOQ) down to 0.1 pM, significantly below the state-of-the-art. We identify the physical parameters setting the LOQ and introduce a broadly applicable figure of merit (FoM) that determines the LOQ and allows for a clear comparison between experimental configurations. Our approach preserves the ability to monitor dynamic interactions, diffusion times, and distinguish species in complex mixtures. We illustrate this feature by measuring the biotin-streptavidin association rate constant which is highly challenging to assess quantitatively due to the strong affinity of the biotin-streptavidin interaction. These findings push the boundaries of single-molecule fluorescence detection for biosensing applications at sub-picomolar concentrations with high accuracy and simplified systems.

## 1. Introduction

Fluorescence techniques have become essential tools to investigate the molecular mechanisms involved in biology, notably thanks to their exceptional sensitivity down to the single molecule level.[1–4] A broad class of fluorescence-based methods is currently available to monitor molecular dynamics with high sensitivity and temporal resolution, including fluorescence time trace analysis,[5] fluorescence correlation spectroscopy (FCS),[6–8] Förster resonance energy transfer (FRET),[9,10] single particle tracking (SPT),[11] or multiparameter fluorescence detection (MFD).[12,13] However, the applications of single molecule fluorescence techniques to biosensing and diagnostics at sub-picomolar concentrations remain scarce [14,15]. Enzyme-linked immunosorbent assay ELISA [16,17] and polymerase chain reaction PCR [18], the two gold standards for protein and DNA detection, do not rely on single-molecule fluorescence detection. As we discuss below, there are many different technical challenges currently preventing the deployment of the rich single-molecule fluorescence toolbox to the quantitative detection of analytes at sub-picomolar concentrations relevant for biosensing.

A confocal microscope equipped with a high numerical aperture (NA) objective is the ubiquitous workhorse for all the different single molecule fluorescence techniques (Fig. 1a). The rationale behind this choice is to maximize the detected fluorescence brightness per molecule while simultaneously minimizing the background from Rayleigh and Raman scattering.[19] However, a direct consequence of the diffraction limit in microscopy is that the detection of single fluorescent molecules is only achievable in a specific concentration range (Fig. 1b) [14,20]. For a 1 fL confocal volume, the concentration to isolate a single molecule is 2 nM. Concentrations significantly deviating from the nanomolar range will result either in ensemble-averaging effects where the single-molecule resolution is lost (high concentration limit, micromolar range) or in a severe degradation of the signal to background ratio (low concentration limit, picomolar range). Many different strategies (involving mainly nanophotonics and plasmonics) have been devoted to extend the high concentration limit and overcome the diffraction limitations in confocal microscopy.[14,20–22] On the contrary, the low concentration limit in the picomolar range remains significantly less documented when it comes to confocal single-molecule fluorescence.[23] The parameters defining the limit of detection (LoD) for confocal microscopy are not clearly stated, nor is the microscope configuration yielding optimum results. Since the infancy of single molecule fluorescence detection, it is commonly believed that the higher the objective NA the better, yet this common belief can be questioned as high NA objectives also lead to low sensing volumes.

Counting the number of molecules in a highly diluted solution resembles "finding a needle in a haystack" problem. For concentrations below 1 picomolar, the average distance between two fluorescent emitters becomes on the order of several micrometers (Supporting Information Fig. S1a), far exceeding the microscope resolution. With such large intermolecular distances, the time between consecutive detection events becomes excessively large, on the order of several seconds in the case of free Brownian diffusion (Fig. S1b). In this scenario, the background contribution can quickly overcome the signal as most of the time the detector accumulates scattered background photons or dark counts.[24,25] To overcome these long waiting times, a possibility involves stirring the sample to accelerate the detection rate.[26] However, a major drawback of this approach is that the burst duration of each single molecule event is also significantly reduced, thereby limiting the amount of collected photons per molecule and leaving the signal-to-background SBR nearly unchanged. Another possibility would be to use a microscope objective with a lower numerical aperture so as to increase the size of the detection volume and detect more easily individual molecules in a diluted solution (Fig. S1c). Here the limitation is that the SBR gets strongly degraded for low NA objectives, with a typical SBR value below 0.5 for NAs lower than 0.4 (Fig. S1d). Lastly, a major challenge which is very often overlooked is that the background level itself also encounters intensity noise fluctuations and can show some temporal correlation. If the background was perfectly known and perfectly constant, it could be subtracted from the detected intensity without much issue. However, because of the Poissonian process of single photon detection, the background level $B$ experiences shot noise fluctuations alike the signal, with a standard deviation equivalent to $\sqrt{B}$. This background noise prevents a basic subtraction of the baseline from the signal to achieve an accurate quantification of the number of molecules at ultralow concentrations.

Here we aim to investigate the limit of quantitation (LOQ also sometimes referred to as the limit of quantification [27]) for single molecule fluorescence counting with fluorescence lifetime correlation spectroscopy (FLCS) at low concentrations in the subpicomolar range. In this work, we define the LOQ as the minimal concentration where FLCS can be used to precisely determine the local number of fluorescent molecules with a relative error equal to 1. Our approach to experimentally characterize the relative errors differs from the classical LOQ definition as the background level plus 9 times the standard deviation.[15] However, our method ensures a comprehensive assessment is provided with all the relevant information available for accurate quantification depending on the end-user needs. This value also differs from the standard definition of the limit of detection (LOD) generally used in biosensing (background level plus 3 times the standard deviation) which assesses the statistical reliability of the measurement against the background.[28–30] We introduce a simple figure of merit (FoM) which is shown to determine the relative error and is broadly applicable to compare between experimental configurations and figure out the best microscope configuration. We also discuss a simple modification of the ubiquitous confocal microscope (introducing a diaphragm on the laser excitation beam) and demonstrate the superior sensing performance achieved using this configuration.

FLCS is a powerful variation of fluorescence correlation spectroscopy (FCS) that uses the lifetime information to improve the measurement accuracy.[31–34] In any single molecule fluorescence experiment, determining the background intensity level is not completely straightforward as this level may fluctuate over time and/or include luminescence emission from contaminants. In FLCS, the known fluorescence lifetime of the target molecule can be used to separate the signal from the background,[32,35–37] allowing for the precise measurement of local concentrations, diffusion coefficients, and biochemical reaction rates. Other applications of FLCS include distinguishing separate species in a complex mixture [38,37,39], and removing afterpulsing artefacts [40]. However, while FLCS has been established as a quantitative method to count single molecules,[31,37,38,41] the factors determining the LOQ in FLCS remain unknown, alike the approaches to improve the LOQ and the definition of a broadly applicable FoM. In this work, we address all these challenges at once. We illustrate the effectiveness of our technique by monitoring the interaction dynamics of biotin binding to streptavidin. The biotin-streptavidin interaction and is a prime example of a biochemical reaction with a very high affinity and has major applications in biotechnology.[42,43] However, its high association rate constant in the range $3.0 \times 10^6$ to $4.5 \times 10^7$ $M^{-1}.s^{-1}$ [44] makes this interaction highly challenging to monitor using standard FCS technique.[45,6] Altogether, our findings open up new possibilities to extend the sensitivity of FLCS in biosensing applications, paving the way for detecting subpicomolar concentrations with high accuracy.

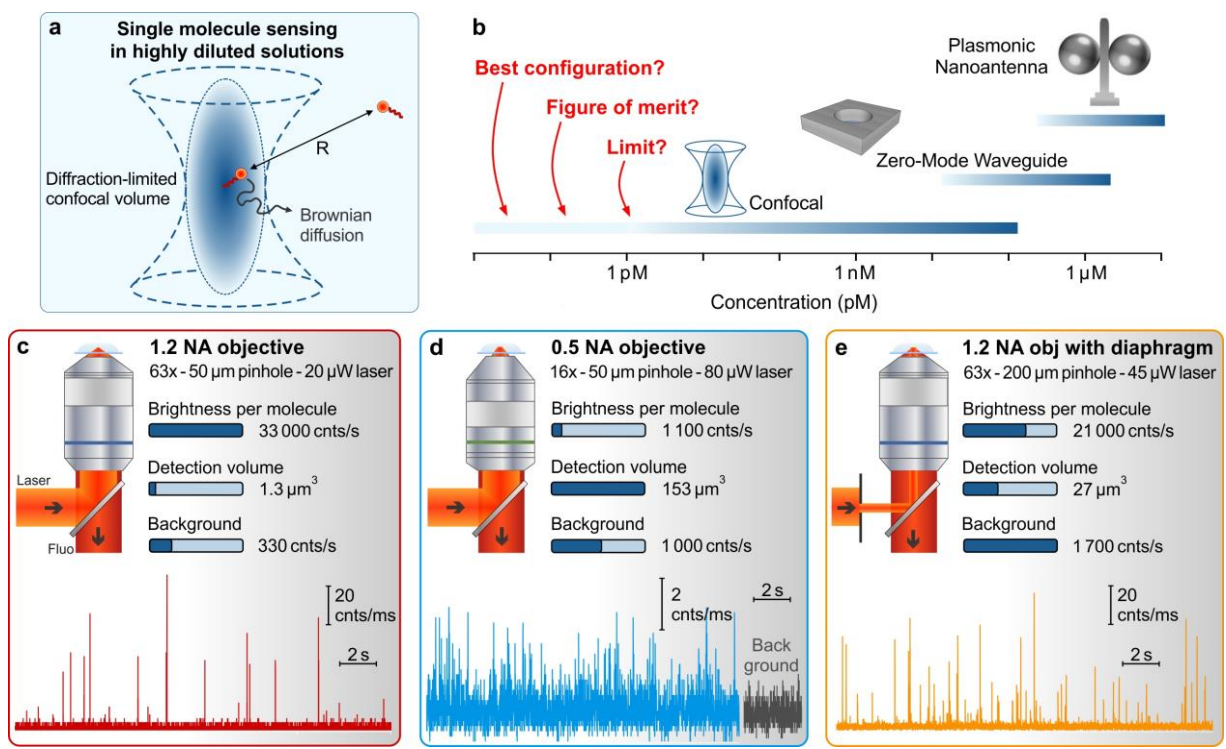

**Figure 1.** Challenges of using confocal microscopy for detecting single molecules at picomolar concentrations. (a) Sketch of the confocal diffraction-limited volume. R is the average distance between individual molecules. (b) Comparison of the practical ranges for single molecule detection between confocal and nanophotonic approaches (including FCS-related methods). (c-e) Comparison between the different confocal microscopes configurations used in this work: (c) high NA microscope objective, (d) low NA objective and (e) high NA objective with a diaphragm on the excitation beam. The time traces at the bottom were recorded experimentally with a 2 pM solution of CF640R fluorescent molecule.

## 2. Results and Discussion

Figure 1c-e summarize our different experimental configurations to explore the limits of FLCS at picomolar concentrations. As fluorescent molecule, we select CF640R for its high brightness, good photostability and relatively long 4.4 ns fluorescence lifetime.[39,46] We also perform our characterization experiments (Fig. 2 and 3) in $D_2O$ buffer to benefit from a 30% increase in brightness thanks to the reduction of the quenching losses induced by water in the red spectral range.[47,48] However, our approach to determine the LOQ and the FoM remains fully general and can be applied to other fluorescent emitters and spectral ranges without problem. While FLCS should be the preferred approach to determine (and correct for) the background on each experimental run, our data will also discuss the use of classical FCS without using the lifetime filtering.

To represent a typical configuration found on confocal microscopes, we use a 1.2 NA water immersion objective (Fig. 1c). This configuration presents the highest fluorescence brightness (count rate per molecule $CRM$) and the lowest background intensity $B$, but also the smallest detection volume. Clear peaks associated with single molecule diffusion events are visible on the fluorescence intensity time trace at 2 pM CF640R concentration (Fig. 1c), but at a relatively low frequency of about 1 event per second. As an alternative for a moderate (lower) NA objective, we use a 0.5 NA water immersion objective (Fig. 1d, quite similar results were obtained while testing a 0.6 NA air objective; they will be not represented here for clarity). The 0.5 NA features the lowest brightness, but allows for the largest detection volume, being 120× larger than the 1.27 fL diffraction limited volume found with the 1.2 NA objective.

The third and last configuration is represented on Fig. 1e: a diaphragm is installed on the laser excitation beam to reduce its diameter to 1.35 mm and underfill the 6.5 mm objective back aperture. The rationale behind this choice is to enlarge the confocal detection volume and voluntarily deviate from the diffraction limit set by the objective's NA. As a consequence of the diaphragm presence, the laser spot size is increased allowing to excite molecules on a larger volume.[49] To account for this, the confocal pinhole has to be increased from 50 to 200 µm. A detailed characterization of the influence of the diaphragm diameter is presented in the Supplementary Information Fig. S2 together with guidelines to set the experimental parameters. Technically, this configuration merges the advantages of the previous two setups (Fig. 1c,d): the excitation is virtually performed as if we were using a low NA objective (large excitation volume), but the collection still takes advantage of the high collection NA (high detection efficiency for high fluorescence brightness). One drawback is that the laser power has to be increased to compensate for the lower excitation intensity due to the larger spot size. This increased power also leads to an increase in the background level, which scales proportional to the volume and the laser power. Currently the brightness $CRM$ for the 1.2 NA with diaphragm case is limited by the maximum power available from our fiber-coupled 635 nm pulsed laser diode (PicoQuant LDH series). We have also calibrated the brightness $CRM$ as a function of the excitation power for all the 3 configurations to ensure our results avoid saturation and are representative of the linear regime of fluorescence.

With the 3 microscope configurations (Fig. 1c-e), we have a set of cases where the count rate per molecule $CRM$ and the detection volume vary by more than 30× and 120× respectively. The questions are now to determine what is the FLCS limit of quantitation (LOQ) for each case, what are the key parameters determining the LOQ and which is the best performing setup. To this end, we perform a series of dilution experiments,[37] starting from concentrations in the nanomolar range and gradually decreasing the CF640R concentration to monitor when the number of molecules $N$ measured with FLCS deviates from the expected linear trend. Figure 2a summarizes our main results. Each experimental data point results from a complete FLCS analysis (see Methods section for details and Supporting Information Fig. S3 for representative raw datasets). Above 10 pM concentrations, the results for all the 3 configurations follow a linear trend with the concentration, where the detection volume determines the slope. Even the 0.5 NA objective allows a quantitative assessment of the number of molecules in this regime, despite the prevailing belief in the field that high NA objectives are essential.

We are primarily interested in noting when the FLCS number of molecules deviates from the linear trend. The deviation is better visualized while plotting the relative error $(N_{measured} - N_{expected})/N_{expected}$ as a function of the concentration (Fig. 2b). Empirically, we find that the relative error follows a power-law dependence on concentration, with a -1.5 exponent across all the three configurations (color lines in Fig. 2b). We have derived a theoretical model (dashed black lines in Fig. 2b) detailed in the Supporting Information Section S6, though the reader may easily omit this demonstration which is not central to the rest of our article. The case with the 1.2 NA objective is the first to deviate at concentrations below 10 pM, followed by the 0.5 NA objective below 2 pM. The configuration using the 1.2 NA objective with diaphragm is the most accurate at low concentrations, with the deviation being only visible at concentrations below 0.5 pM (Fig. 2a,b).

As briefly mentioned in the introduction, two different definitions come into play while discussing the limit of detection. Generally in biosensing, the limit of detection LOD is defined as the minimal concentration where the signal exceeds the background plus three times the standard deviation.[29,30] This LOD definition ensures the reliability of the measurement against the background ensuring that the probability of a false negative is typically below 1%. However, determining the presence or absence of fluorescent molecules against a certain background is generally not the question of interest for FCS users. Instead, the question focuses on precisely determining the local number of fluorescent molecules with a chosen level of accuracy. Determining the maximum acceptable error *a priori* depends on the targeted application. Here we arbitrarily consider the concentration where the relative error in $N$ amounts to 1, and use this value to define the so-called limit of quantitation LOQ (also known as limit of quantification). The LOQ is more demanding and generally higher than the LOD, which is expected to be found in the tens of femtomolar range (see a discussion in Fig. S4 in the Supplementary Information together with background correlation data).

Our approach to define the LOQ depends on the level of accuracy desired, there is an element of arbitrariness in this decision.[27,50] The curves displaying the relative errors in $N$ as a function of the concentration provide all the information and can readily be used to define which concentration gives a maximum tolerable relative error (Fig. 2b). Choosing this maximum acceptable error depends on the targeted application. Here we arbitrarily decide to define the LOQ as the concentration where the relative error in $N$ amounts to 1. For the 1.2 NA objective, our results indicate a LOQ of 0.75 ± 0.05 pM, going down to 0.3 ± 0.1 pM for the 0.5 NA objective and 0.15 ± 0.05 pM for the 1.2 NA objective with diaphragm. These results clearly indicate the superior sensing performance of the configuration using the 1.2 NA objective with diaphragm. Achieving an intermediate detection volume of a few tens of femtoliter while still preserving the fluorescence brightness per molecule is the key to improve the FLCS accuracy at sub-picomolar concentrations and push its sensitivity.

To further validate our approach, we compare our results with the IUPAC definition of the limit of quantification (LOQ), which sets the threshold as the average background signal plus nine times its standard deviation. Repeated experiments on the blank solution in absence of fluorescent target molecules assess the average background and its standard deviation (Fig. S4a,b). Considering the mean background plus nine times the standard deviation results in a detection limit in the 10 fM range (Fig. S4c,d), an order of magnitude lower than our reported values. This finding ensures that the false positive and false negative rates remain well below 1%,[15] highlighting the conservative nature of our claims and reinforcing the robustness of our approach. This analysis is consistent with the discussion identifying systematic errors, likely stemming from residual shot noise fluctuations, as key parameter defining the FLCS accuracy (Fig. 3).

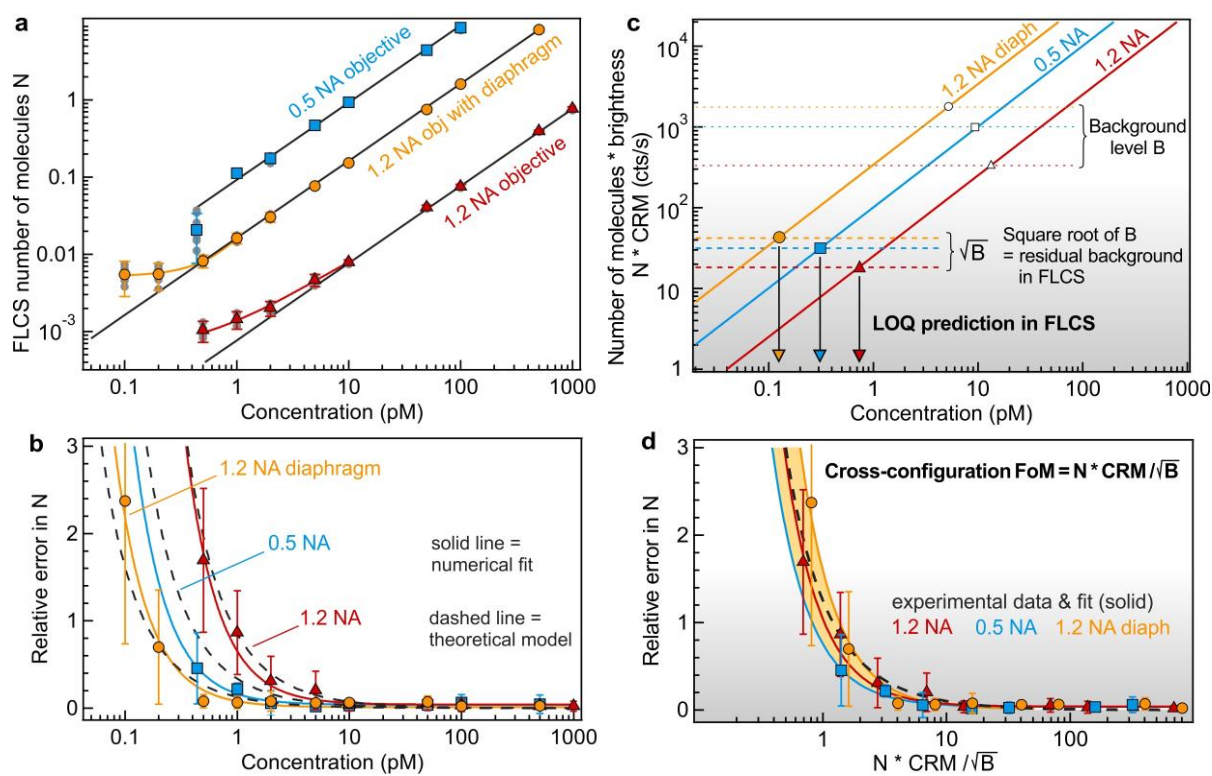

**Figure 2.** Determining the best performing microscope configuration and the limit of detection for FLCS at low concentrations. (a) Evolution of the average number of molecules measured by FLCS as function of the CF640R concentration. Throughout this figure, markers are experimental data, and the error bars represent twice the standard deviation (number of experiments and integration times are detailed in the Supporting Information Tab. S1). The black lines are the expected values based on the detection volumes measured at concentrations above 10 pM (serving as ground truth information here). When the experimental data deviates significantly from the black line prediction, the FLCS method becomes inaccurate. (b) Absolute value of the relative error in the FLCS number of molecules as function of the concentration for the different configurations, determined from the data in (a). The solid color lines are a numerical fit to the data, while the dashed lines correspond to a theoretical model based on Eq. (S4). The configuration with the 1.2NA objective with diaphragm is the most accurate (lowest relative error) for low concentration detection. (c) Product of the number of molecules in the detection volume $N$ times their fluorescence brightness $CRM$ as function of the concentration for the different configurations in Fig. 2. The points where this quantity corresponds to the background level $B$ and the square root of $B$ are indicated by markers. As seen in (a,b), FLCS is still accurate when $N * CRM < B$ (empty markers, dotted lines) but the relative error increases dramatically when $N * CRM < \sqrt{B}$ (filled markers, dashed lines). (d) Relative error in the FLCS number of molecules (similar to a) as function of the quantity $N * CRM/\sqrt{B}$. When the errors are plotted as function of this term, all the three cases collapse (within experimental uncertainties) into a single trend. This observation demonstrates that the quantity $N * CRM/\sqrt{B}$ is the key factor defining the accuracy in FLCS detection of the number of molecules, leading us to use it for our definition of the FoM.

While the data in Fig. 2a,b indicate when the FLCS deviates from the expected trend and loses is accuracy, the observations do not elaborate on the origins of this effect and what are the key parameters determining the LOQ. We have performed different sets of experiments to further explore the origin of the accuracy deviation. First, theoretical calculations indicate that the deviation is not a statistical problem, as the signal-to-noise ratio (SNR) is always greater than 1 for all the 3 microscope configurations at concentrations above 0.05 pM (Supporting Information Fig. S5). This result is further confirmed by the fact that increasing the FLCS integration time from 3 to 18 minutes does not affect the relative error (Fig. S6). From these observations, we can conclude that the accuracy deviation is a systematic error and is not related to some statistical noise affecting the measurement. In a second set of experiments, we have stirred the solution containing the CF640R molecules while performing

FLCS (Fig. S7). The idea is to speed up the diffusion process so as to increase the frequency at which fluorescence events are detected. While the diffusion time is accelerated by more than 10×, again the relative error remains unchanged. This result confirms the error invariance with the integration time: acquiring more single-molecule bursts surprisingly does not improve the accuracy at subpicomolar concentrations. Lastly, applying time-gating [36] enables an improvement in the accuracy (Fig. S8). Here photons are rejected based on their arrival time before the FLCS filters are applied, allowing to virtually reduce the background level $B$. The selection of tighter time-gated intervals reduces the systematic error (Fig. S8). This indicates that the background level $B$ is a key factor determining the FLCS deviation.

To better visualize this effect, we plot in Fig. 2c the product of the expected number of molecules $N$ times their fluorescence brightness $CRM$ as function of the concentration. This quantity $N * CRM$ represents the net fluorescence signal collected form the detection volume. For each microscope configuration, we indicate the background intensity level $B$ on the same graph and highlight the concentration where the signal-to-background SBR goes below 1 (white markers in Fig. 2c). This typically occurs for concentrations around 5 to 10 pM for the different microscopes, yet FLCS remains accurate for concentrations at least one order of magnitude lower. As the core concept of FLCS is to filter out the photons using the lifetime information to distinguish between the signal and the background, it is thus not surprising that FLCS remains accurate at conditions where SBR < 1. More intriguing is the observation that the LoDs as defined on Fig. 2b correspond to the levels where the net signal $N * CRM$ amounts to $\sqrt{B}$ for all the three configurations (filled markers in Fig. 2c). To elaborate further on this observation, we reconsider the relative errors in FLCS (similar to Fig. 2b) and plot them as function of the quantity $N * CRM/\sqrt{B}$ (Fig. 2d). The errors for all the three setups then collapse into a unique trend (within experimental uncertainties), although the configurations differ by more than 100× in detection volumes and 30× in molecular brightness. This demonstrates that the quantity $N * CRM/\sqrt{B}$ is the key factor defining the accuracy in the FLCS detection of the number of molecules.

We thus define the figure of merit in FLCS as FoM = $N * CRM/\sqrt{B}$. The presence of a term in $\sqrt{B}$ instead of the background level $B$ can be explained as a residual contribution from the shot noise on the background level that remains uncorrected by FLCS. While FLCS is very efficient in filtering signal photons from the background for each experimental run, it does not incorporate the fact that due to the Poissonian nature of the detection event (being signal or background), there is inevitably a shot noise contribution whose uncertainty scales as the square root of the average intensity.[51] Hence if the total count rate per second is $I_0$, the standard deviation of the shot noise will be $\sqrt{I_0}$. We have verified experimentally that this reasoning still holds for the background level: the standard deviation of the noise on the background $B$ indeed scales as $\sqrt{B}$ (Supporting Information Fig. S9 and Tab. S2) meaning that shot noise is the dominant source of noise of our system and that electrical or computer noise remain negligible.

Having a figure of merit for FLCS allows to directly compare between configurations and even predict the LOQ. When the FoM becomes lower than 1, the relative error increases strongly and the FLCS technique becomes inaccurate (see Fig. S10 for a plot of this FoM as function of the concentration). In the conditions used here where the SNR is not an issue, the FoM does not depend on the molecular diffusion time, nor on the number of events per seconds nor on the total integration time (Fig. S5-S8). Achieving the lowest LOQ concentration boils down to maximizing the detection volume and the $CRM$ brightness while minimizing the background level. For this purpose, the configuration in Fig. 1e with the 1.2 NA objective and the diaphragm achieves the best performance and highest FoM (Fig. S10), with nearly one order of magnitude gain in FoM as compared to the classical 1.2 NA case. Even more surprising is the observation that the 0.5 NA objective yields a 2.5× better FoM than the classical 1.2 NA configuration, challenging the usual belief in FCS that the highest NA objective always yields the best results. Sensing at low concentrations outside the usual FCS and FLCS ranges imposes to rethink this postulate.

In regular FCS, the influence from a non-negligible background is often accounted for by introducing a supplementary correction factor $\left(1 - \frac{B}{B+N*CRM}\right)^2$.[6,52] We now investigate if a similar correction could be applied to FLCS as well. Having demonstrated the superior sensing performance of the 1.2 NA objective with diaphragm, we consider this case for the data in Fig. 3 (similar results can be obtained with the 1.2 NA objective without diaphragm, see Fig. S11). Since the residual background contribution goes with $\sqrt{B}$ in FLCS, we introduce a correction factor $\left(1 - \frac{\sqrt{B}}{\sqrt{B}+N*CRM}\right)^2$ for FLCS, where the $B$ term in the classical FCS correction has been replaced by $\sqrt{B}$ (Fig. S12). Figure 3a presents our experimental results and compares between the four different cases of FCS or FLCS with and without correction. Without surprise, regular FCS without background correction is the least accurate method, and deviates from the predicted values already at concentrations around 100 pM. Introducing the background correction term in FCS improves the situation, yet the difficulty to precisely measure the background level $B$ for each experimental set and the variability of $B$ add systematic errors. As a consequence, the data from FCS with correction deviate at concentrations below 10 pM. As we have seen in Fig. 2a, uncorrected FLCS performs better than background-corrected FCS, though the deviation starts at around 0.5 pM. Finally the $\left(1 - \frac{\sqrt{B}}{\sqrt{B}+N*CRM}\right)^2$ background correction in FLCS further improves the accuracy by approximately 3×, pushing the LoD below 0.1 pM. To the best of our knowledge, this is the first time that a FCS-related technique is proven accurate down to concentrations of 100 fM. With the molecular mass of CF640R being only 832 Da, this LOQ corresponds to ~100 fg/mL. Importantly the detection is not limited to the quantification of the number of molecules, the key FCS abilities to analyze the diffusion time and distinguish between species in a mixture are still preserved.

Figure 3b compares the relative error in the FLCS number of molecules at 0.2 pM concentration after the application different corrections. As discussed in Fig. S6 and S7, increasing the integration time and accelerating the diffusion to detect more bursts per second have negligible effects on the relative error. Time gating (Fig. S8) and most importantly FLCS background correction reduce the relative error. The evolution of the relative errors for the different methods is given in Fig. 3c as function of the concentration, allowing to determine the respective LOQ for each technique as in Fig. 2b. Alternatively, we can compute the accuracy gain respective to standard FCS by taking the ratio of the relative errors for standard FCS and the technique of interest (Fig. 3d). The absolute value of the accuracy gain can reach huge values up to 600×, yet this is primarily because error for the standard FCS becomes very large. The main interest and message from Fig. 3d is that the FLCS background correction outperforms the other approaches (uncorrected FLCS and background-corrected FCS) in terms of accuracy gain. This highlights the interest for taking into account a supplementary correction for the $\sqrt{B}$ background fluctuations while performing FLCS at ultralow concentrations.

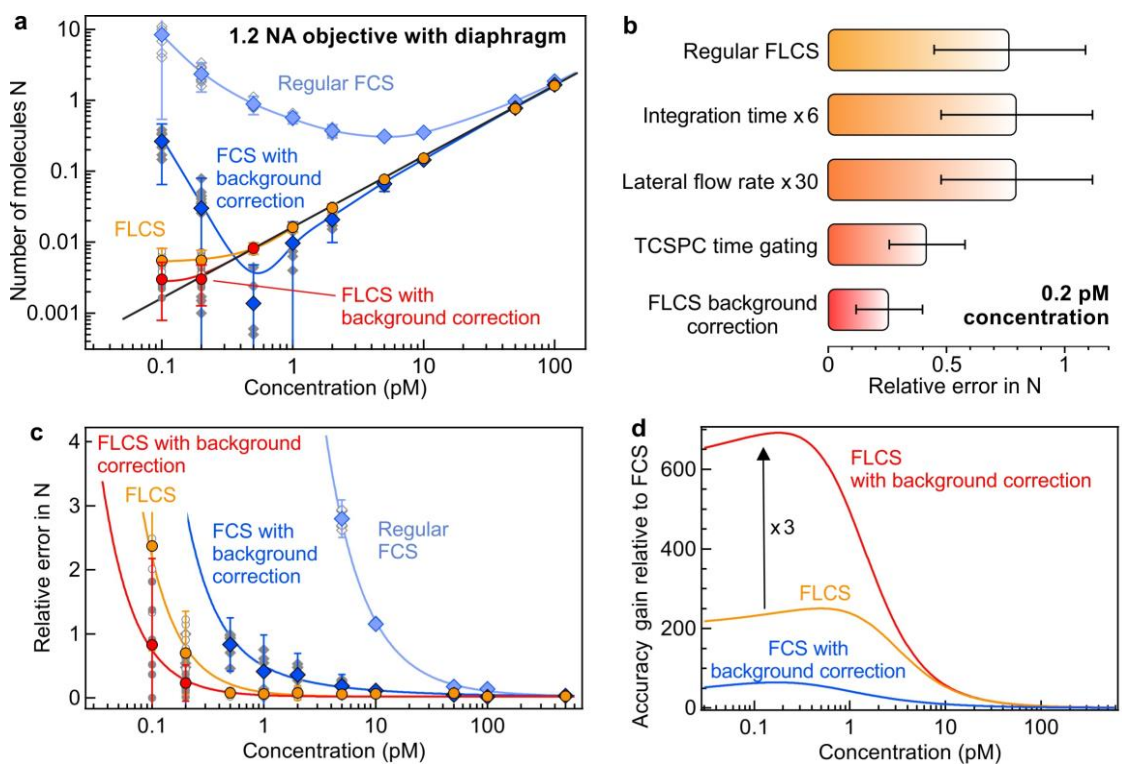

**Figure 3.** Background correction in FLCS to improve the accuracy at low concentrations. All these data were recorded on the 1.2NA objective with diaphragm, as this configuration was determined in Fig. 2

to provide a superior performance. (a) Evolution of the average number of molecules measured with different techniques (FCS and FLCS, with and without background correction, see text for details) as function of the CF640R concentration. Markers are experimental data, error bars correspond to twice the standard deviation. The number of experiments and integration times are detailed in the Supporting Information Tab. S1. The black line is the expected trend (ground truth) based on the detection volumes measured by FLCS at concentrations above 10 pM. The color lines are a guide to the eye to better visualize the deviation from the linear prediction. (b) Evolution of the relative error at 0.2 pM concentration when different modifications are applied to the experimental protocol. (c) Absolute value of the relative error in the number of molecules as function of the concentration for the different configurations, determined from the data in (a). Lines are a numerical fit to the data. (d) Accuracy gain as compared to regular (uncorrected) FCS as function of the concentration, determined from the ratios of the relative errors in (c).

To demonstrate the applicability of the technique beyond accurate concentration measurements in the (sub)picomolar range, we investigate the association dynamics between biotin and streptavidin (Fig. 4a). This association is largely used in biotechnology for its high affinity, with its dissociation constant $K_d$ around $4.10^{-14}$ M [53] making it one of the strongest noncovalent interactions in biology.[42,43] However, despite its major role, the measurement of the biotin-streptavidin association rates remains scarce.[44,46,54–56] While the dissociation rate constant is very slow and practical to measure,[57–59] the fast association rate around $10^7$ M$^{-1}$.s$^{-1}$ makes the binding challenging to monitor using conventional methods.[45,54] Microfluidics [44,56] and silicon nanowires [55] were used to probe the association dynamics at nanomolar concentrations, while nanophotonics was needed to achieve short enough FCS integration times to probe the interaction [46].

Here we take advantage of our high sensitivity at low concentrations to work in the picomolar range where the binding times can be assessed by FLCS. In our experiments, biotin labelled with Atto643 is mixed with label-free streptavidin. The biotin-Atto643 concentration is kept constant at 10 pM concentration, while the streptavidin concentration varies between 50 and 200 pM (Fig. 4a). Under these conditions, the association rate is controlled by the streptavidin concentration as the biotin concentration remains low enough. Moreover, we can assume that each single biotin binds to a different streptavidin, so the fact that each streptavidin holds 4 biotin binding sites can be neglected. The microscope uses the configuration with the 1.2 NA objective with diaphragm to benefit from a better accuracy as demonstrated in Fig. 2 & 3. The average number of fluorescent molecules remains around 0.1 (Fig. S13) so that single molecule conditions can be claimed, as exemplified on the fluorescence time traces showing clear single-molecule fluorescence bursts (Fig. 4b).

To separate between the free and bound fractions of biotin, we perform a FLCS analysis with two species, taking advantage of the longer diffusion time of the streptavidin-bound biotin respective to the free biotin (see Methods Section for details). Upon binding to streptavidin, the Atto643-biotin diffusion time gradually evolves from 0.9 ms (free biotin) to 3.7 ms (fully bound biotin) (Fig. 4c). This 4.1× increase in the diffusion time is consistent with the change of molecular weight from 1.26 kDa (free Atto643-biotin) to 67 kDa (Atto643-biotin-streptavidin complex) as $\sqrt[3]{67/1.26} \sim 3.8$.[45] It is also well beyond the 1.6× minimum difference in diffusion times needed in FCS to distinguish 2 components.[60]

The association dynamics are presented on Fig. 4d,e for different streptavidin concentrations. The bound fraction increases exponentially with a characteristic time $\tau = 1/(k_{on}[S] + k_{off})$, where $k_{on}$ and $k_{off}$ are the association and dissociation rate constants and $[S]$ is the streptavidin concentration. Finally, the plot of $1/\tau$ as a function of $[S]$ allows to determine the values for the rate constants $k_{on}$ and $k_{off}$ (Fig. 4f). The slope determines the association rate constant $k_{on}$ to be (2.7 ± 0.2) × 10$^7$ M$^{-1}$.s$^{-1}$ while the intercept at origin provides an upper bound for the dissociation rate constant $k_{off}$ to be well below 10$^{-4}$ s$^{-1}$. These values agree well with the data reported in the literature, more details are provided in the Supporting Information Table S3. The key aspect of this application is to demonstrate the capability of conducting FLCS studies to distinguish two species for a prime example of a fast association rate, enabled by the extended operational range down to picomolar concentrations.

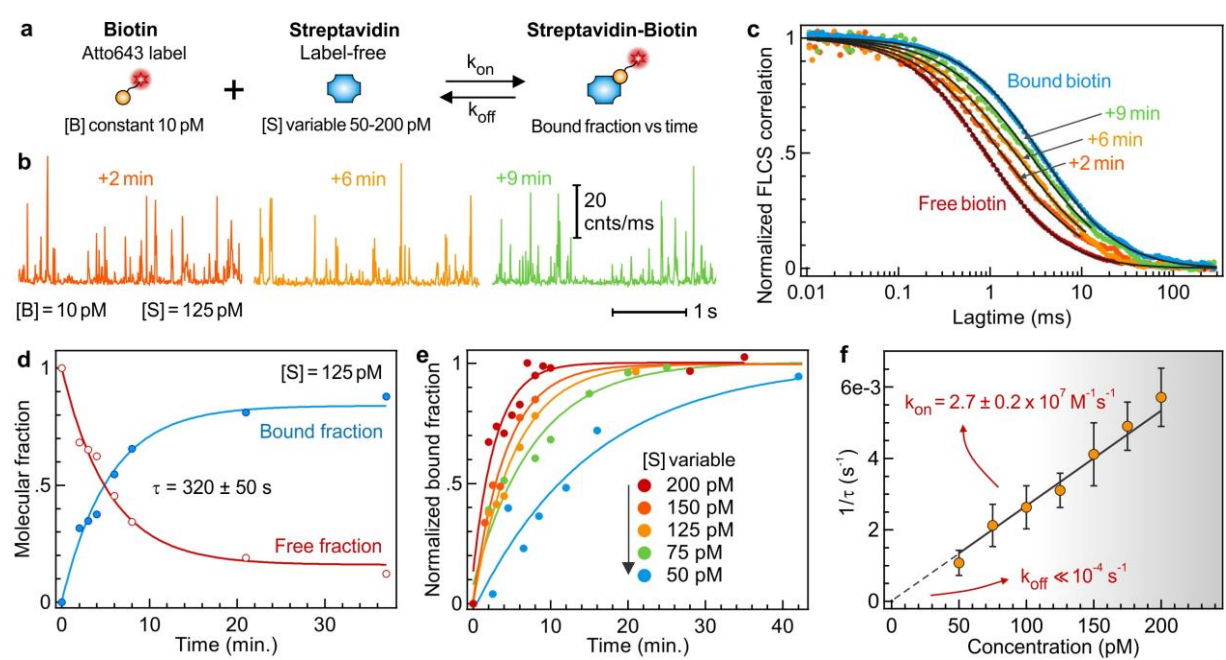

**Figure 4.** Application of FLCS sensing at low concentrations to determine biotin-streptavidin association dynamics. (a) Sketch of the experiment: biotin labeled with Atto643 is mixed with unlabeled streptavidin to determine the bound fraction of biotin-streptavidin over the reaction time. Due to the high affinity of the reaction and the fast association rate, picomolar concentrations are needed to enable monitoring the reaction dynamics with single molecule sensitivity. (b) Time traces demonstrating the detection of single molecule fluorescence bursts at 10 pM of labeled biotin. The burst duration increases over the time of the experiment due to the addition of 125 pM of streptavidin, yet FLCS is better suited to analyze the binding. (c) Normalized FLCS correlation curves of Atto643-biotin mixed with 125 pM streptavidin at T = 0. The shift towards longer diffusion times indicates binding of biotin to streptavidin. (d) Temporal evolution of the free and bound molecular fractions of biotin after adding 125 pM solution of streptavidin. Markers are experimental data, lines are exponential fits with characteristic time $\tau$. (e) Comparison of the temporal evolutions of the bound fractions with decreasing streptavidin concentrations. (f) Inverse of the association time $\tau$ as a function of the streptavidin concentration. Markers indicate the average value, the error bars represent the fit uncertainty upon processing the data in d,e with Igor Pro 7 (Wavemetrics) . The slope determines the association rate constant $k_{on}$ while the interpolation of the intercept at origin gives an upper bound for the dissociation rate constant $k_{off}$.

## 3. Conclusions

It was largely assumed that single molecule fluorescence techniques and FCS were only functional at concentrations above 50 pM, and that detection at lower concentrations was limited by a severe loss in signal to background ratio.[19] This so-called low concentration barrier was identified as one of the main limitations preventing the application of the rich single molecule fluorescence toolbox to the biosensing applications in biology and medicine.[14] Here we show that the low concentration limit can be overcome with a limit of quantitation LOQ down to 0.1 pM without any major modification to the single molecule confocal microscope and without the introduction of complex microfluidics or preconcentration stages. By adding a simple diaphragm on the laser excitation beam, our microscope achieves simultaneously a large detection volume of a few tens of femtoliter and a high fluorescence brightness per molecule. The combination of these two factors enables a 8× superior sensing performance than the classical confocal microscope and is the key to push the FLCS sensitivity towards sub-picomolar concentrations.

For the first time, we clearly discuss the physical parameters setting the lower the limit of quantitation for FLCS and we introduce a universal figure of merit $N * CRM/\sqrt{B}$ allowing to compare between experimental configurations and predict the LOQ. Our results establish that the LOQ in FLCS is not a statistical problem determined by the noise on the signal or an issue related to the low frequency of the detection events. Instead, we identify the residual shot noise fluctuations on the background level as a key parameter determining the LOQ. While standard FLCS does not account for this effect, we show that it can be at least partly compensated by introducing a post-measurement correction factor in the form $\left(1 - \frac{\sqrt{B}}{\sqrt{B} + N*CRM}\right)^2$.

Importantly our results are not limited to the measurement of the local concentration, the key abilities to analyze the diffusion time and distinguish between species in a mixture remain preserved, as we demonstrate by monitoring the interaction dynamics of biotin binding to streptavidin. The extended operational range down to picomolar concentrations opens new opportunities to measure fast association rate constants while keeping regular single molecule fluorescence resolutions.

The current LOQ of 0.1 pM is found for FLCS with standard commercial optics and fluorescent dyes. Obtaining brighter probes or reducing further the background level would immediately impact the LOQ in a manner that can be predicted using our equations. It should also be stressed that this LOQ applies for FLCS, allowing to analyse local concentrations, brightness, translational and rotational diffusion as well as distinguishing subpopulations in a mixture.[33,34] If only fluorescence bursts are to be measured, then the LOD is expected to be found in the femtomolar range, yet at the expense of a lower amount of information.[24,26] Complementary approaches using micro/nanofluidics or preconcentration could be used as well to further improve the sensitivity to femto or attomolar ranges.[61–63] In conclusion, our work opens up new directions to extend the sensitivity of FLCS and overcome the low-concentration barrier in single molecule studies while keeping simple instrumentation. This paves the way for future biosensing applications at sub-picomolar levels with single molecule resolution.

**Experimental Section**

**Sample Preparation**

CF640R and solvent D$_2$O is purchased from Sigma-Aldrich and used as received. Streptavidin extracted from Streptomyces avidinii was purchased from Sigma Aldrich and it was stored in -20°C. Biotin tagged with Atto 643 (Biotin-Atto643) dyes were purchased from ATTO-TEC. For measurements, Streptavidin and Biotin-Atto643 samples were diluted in phosphate-buffered saline (PBS) solution. Biotin-Atto643 in PBS was sonicated for 20-30 min before every experiment to prevent aggregation and form a uniform sample.

To prevent adsorption of streptavidin, the coverslip surface was passivated using polyethylene glycol (PEG) silane (Nanocs PG1-SL-IK) of 1000 Da molar mass following the protocol in ref.[64]. Before passivation, the coverslip was cleaned with water and ethanol (96%). Further the coverslip was dipped in isopropanol and sonicated to 10 min to remove any dirt from the surface. After this, the sample was exposed to UV ozone for 10 min to remove organic impurities. Finally the samples were put in air plasma cleaner for 10 min, after which the sample was immediately transferred to a sample holder and covered with 1 mM PEG solution prepared on 96% ethanol and 1% acetic acid. The chamber was blown with argon gas and left at room temperature for overnight. The next day sample was rinsed with ethanol and dried using nitrogen. The passivated sample was then stored in a 1% tween20 solution in 96% alcohol and stored in a fridge.

**Confocal microscope setup**

All the FCS measurements are done in a home-built confocal microscope (Nikon Ti-U Eclipse) equipped with 635 nm pulsed laser (LDH series laser diode, PicoQuant) with a pulse duration $\approx 50 \text{ ps}$. To reflect the laser towards the microscope we used a multiband dichroic mirror (ZT 405/488/561/640rpc, Chroma). Further, Zeiss 63x, 1.2 NA water immersion or Zeiss 16x, 0.5 NA water immersion are used to focus the excitation light as well as collect the emission in the epifluorescence configuration. For the use of the diaphragm (case Fig. 2c), a diaphragm of 1.35 mm diameter corresponding to a 35.4% transmission is used to reduce the laser beam diameter and underfill the objective's back aperture (measuring the transmission through the diaphragm is more convenient than measuring its diameter, it is the recommended method to ensure an accurate reproducibility of the data collection). For the rejection of backscattered laser the emission is passed through the same multiband dichroic mirror (ZT 405/488/561/640rpc, Chroma) and then through two emission filters (ZET405/488/565/640mv2 and ET655, Chroma). A 50 µm (200 µm for diaphragm case) pinhole is used to spatially filer the molecular fluorescence. The signal is collected by a single-photon avalanche photodiode (APD) (Perkin Elmer SPCM AQR 13) in the range 650-750 nm. In order to block the entering of stray light in the APD and for the rejection of Rayleigh scattered light we used a supplementary emission filter (FF01-676/37, Semrock), placed in front of the APD. The photodiode output is connected to a time-correlated single photon counting (TCSPC) module (HydraHarp 400, Picoquant). Table S1 in the Supporting Information summarizes the number of experiments and integration times for the different configurations. The background intensity $B$ was recorded using pure D$_2$O buffer in the same conditions as for CF640R dyes.

**FLCS Analysis**

Symphotime64 (Picoquant) is used for the FLCS analysis. The lifetime decay, $D(i)$ of the compound is composed of photons from the sample and background contribution from the scattered excitation light, detector dark counts, detector afterpulsing and residual room light. At low concentrations (pM range) the background contribution is not negligible. The stray room light, dark counts and afterpulses are random events. Hence its TCPC pattern is a flat line on 25 ns histogram time scale. The scattered photons forms a bump at the beginning of the TCSPC decay curve. Therefore $D(i)$ can be expressed as a linear combination of 3 patterns; pure fluorescence, scattering and other residual background contribution. Thus $D(i)$ is given by Eq. (1)

$$D(i) = \omega_F\, d_F(i) + \omega_S d_S(i) + \omega_B d_B(i) \tag{1}$$

where $\omega_F$, $\omega_S$ and $\omega_B$ are the photon count amplitude (in number of photons) of the fluorescence signal, the scattering and residual background respectively. $d$ is the normalized TCSPC decay pattern. Here the index $i$ is the channel number for the photons detected within the TCSPC channel. Then a statistical filter function, $f(i)$ is introduced that can satisfy the following requirements.

$$\sum f(i) \times D(i) = \omega \tag{2}$$

$$f^F + f^S + f^B = 1 \tag{3}$$

These filter functions can be calculated numerically based on the TCSPC histogram. Here, we used a pattern-matching filter to calculate the statistical filter function for the sample and background contributions. In this method, we calculate the statistical filter function by appropriately scaling the decay curve of the sample using the decay pattern obtained for the sample at a higher concentration (nM range) with negligible background and the decay pattern of the background recorded in the absence of fluorescent molecules. This gives us 3 filter functions; fluorophore, scattered photons and residual flat background. Once we have the filter functions, the filtered correlation can be calculated using Equation (4)

$$G(\tau) = \frac{\langle \sum_i f_i I_i(t) \times \sum_i f_i I_i(t+\tau) \rangle}{\langle \sum_i f_i I_i(t) \rangle^2} \tag{4}$$

This correlation function is fitted using a pure diffusion fitting model given by

$$G(\tau) = \frac{1}{N}\left(1 + \frac{\tau}{\tau_D}\right)^{-1} \times \left(1 + \frac{\tau}{\tau_D \kappa^2}\right)^{-1/2} \tag{5}$$

Here $G(\tau)$ is the autocorrelation function at time $\tau$, $N$ is the total number of CF640R molecules in the observation volume, $\tau_D$ is the mean diffusion time and $\kappa$ denotes the aspect ratio of the axial to the transversal dimension of the detection volume. In our case, $\kappa$ is taken as 5 based on the confocal measurements done in the past and fits well with our experimental correlation data.

**Single Molecule Burst Analysis**

For this work, we used the Burst Analysis Module of the PIE Analysis with MATLAB (PAM).[65] All Photon Burst Search function is used for burst detection by putting the minimum photons per burst, the time window of the bursts, and the total photon counts per burst. These values differ for three cases considered 1.2 NA, 0.5 NA, and 1.2 NA with diaphragm due to the difference in diffusion time and background signal. For 1.2 NA (0.5 NA/1.2 NA with diaphragm) configuration, each peak was considered a single molecule burst having at least 30 (30/30) photons. We have taken a minimum of 2 (2/2) photons per time window of 100 (1000/600) µs.

**Streptavidin-Biotin association rate calculation**

The association and dissociation dynamics of streptavidin and biotin can be denoted as

$$S + B \underset{k_{off}}{\overset{k_{on}}{\rightleftharpoons}} SB$$

Here S denotes free streptavidin, B is free biotin and SB is Streptavidin-Biotin complex. Streptavidin has four biotin binding sites. For our measurements we have taken the biotin concentration to be significantly lower than that of streptavidin. Hence we can safely assume that only a single biotin binds to single streptavidin making the kinetic study easier compared to multiple binding to the streptavidin. $k_{on}$ is the association rate constant to form the complex SB while $k_{off}$ is the dissociation rate constant. The time origin t=0 is set when streptavidin is added to the biotin solution. Differential equations to study the reaction kinetics can be written as:

$$\frac{d[B]}{dt} = -k_{on}[S][B] + k_{off}([B_{tot}] - [B]) \qquad (6)$$

$$\frac{d[SB]}{dt} = k_{on}[S][B_{tot}] - (k_{on}[S] + k_{off})[SB]) \qquad (7)$$

The total biotin concentration is constant and it can be written as $[B_{tot}] = [B](t) + [SB](t)$. Applying the boundary condition that $[B](0) = [B_{tot}]$ and $[SB](0) = 0$ we can solve the differential equations (6) $and$ (7) yielding us,

$$\frac{[B](t)}{[B_{tot}]} = e^{-(k_{on}[S]+k_{off})t} + \frac{k_{off}}{k_{on}[S]+k_{off}}\left(1 - e^{-(k_{on}[S]+k_{off})t}\right) \qquad (8)$$

Dissociation rate constant $k_{off}$ is much lower than $k_{on}[S]$. Hence we can easily neglect the second term in the right hand side of eqn. (8).

The bound fraction is:

$$\frac{[SB](t)}{[B_{tot}]} = \frac{k_{on}[S]}{k_{on}[S] + k_{off}} \left(1 - e^{-(k_{on}[S]+k_{off})t}\right) \quad (9)$$

Considering a characteristic time $\tau = 1/(k_{on}[S] + k_{off})$ we can express free and bound fraction of biotin as an exponential temporal evolution $e^{-t/\tau}$.

We use pure diffusion model with two species to fit the FLCS curve obtained after FLCS filtering to determine the free and bound ratio of biotin. We set the diffusion times of biotin and Streptavidin-Biotin complexes to be 0. 9 ms and 3.7 ms respectively. These values were obtained by independent measurements at nM concentrations. FLCS correlation amplitude $\sigma_1$ and $\sigma_2$ corresponds to free and bound biotin respectively based on diffusion time. The free and bound fractions can be written as:

$$\frac{\sigma_1}{\sigma_1 + \sigma_2} = \frac{[B](t)}{[B_{tot}]} \quad (10)$$

$$\frac{\sigma_2}{\sigma_1 + \sigma_2} = \frac{[SB](t)}{[B_{tot}]} \quad (11)$$

These ratios can be fitted using exponential functions as follows

$$\frac{\sigma_1}{\sigma_1 + \sigma_2} = 1 - a\left[1 - e^{(-t/\tau)}\right] \quad (12)$$

and

$$\frac{\sigma_2}{\sigma_1 + \sigma_2} = a\left[1 - e^{(-(t-t_0)/\tau)}\right] \quad (13)$$

Here the term $t_0$ is to correct for any delay during the measurement. It the tagging efficiency of biotin to Atto 643 is 100 %, the value of a is 1. But we observed a value of ~ 0.8 in our case, meaning tagging efficiency is ~ 80%. We obtain the association time τ by fitting the free and bound ratio of biotin. The value of association rate constant $k_{on}$ is obtained from the slope of $1/\tau$ vs [S].

**Statistical Analysis**

Pre-processing: the intensity time trace was filtered by FLCS to compute the lifetime-specific correlation function, as described in the FLCS Analysis section. No other processing nor filtering was applied to the data. The data presentation correspond to the average, the error bars represent 2 times the standard deviation. The sample size for each statistical analysis is detailed in the Supplementary Information Tab. S1. No statistical test was applied, as we are not comparing different populations or samples. The statistical analysis was performed using Symphotime64 (Picoquant) and IgorPro 7 (Wavemetrics).

**Supporting Information**

Challenges of detecting single molecules at low concentrations; Characterization of the influence of the diaphragm diameter; Examples of FLCS filters and raw correlation data; Number of experiments and integration times for the different configurations; Background correlation in absence of target molecule; Interpolation of the relative error; Signal to noise ratio in FCS; Influence of the integration time on the FLCS error; Influence of the lateral flow on the FLCS error; Influence of time gating on the FLCS error; Shot noise dependence of the background noise; FLCS figure of merit as function of concentration; FLCS background correction applied to the 1.2NA microscope configuration; Background correction prefactors for FCS and FLCS; Number of biotin molecules and brightness per molecule as function of streptavidin concentration; Literature values for streptavidin-biotin binding kinetics.

**Acknowledgments**

This project has received funding from the European Research Executive Agency (REA) under the Marie Skłodowska-Curie Actions doctoral network program (grant agreement No 101072818).

**Conflict of Interest**

The authors declare no conflict of interest.

**Data Availability Statement**

The data that support the findings of this study data are available from the corresponding author upon request.

**Supporting Information for**

**Breaking the Low Concentration Barrier of Single-Molecule Fluorescence Quantification to the Sub-Picomolar Range**

Malavika Kayyil Veedu,[1] Jérôme Wenger[1,*]

[1] *Aix Marseille Univ, CNRS, Centrale Med, Institut Fresnel, AMUTech, 13013 Marseille, France*

*\* Corresponding author: jerome.wenger@fresnel.fr*

**Contents:**

## S1. Challenges of detecting single molecules at low concentrations

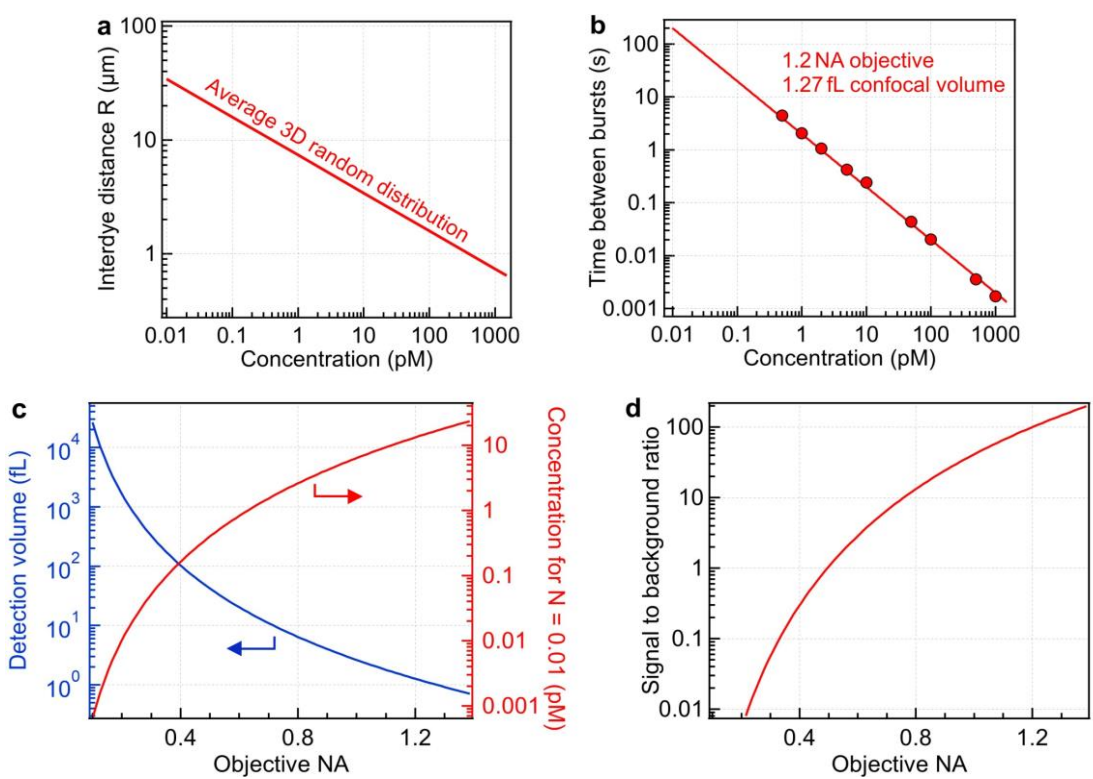

**Figure S1.** (a) Evolution of R as a function of the concentration, assuming a random distribution of the molecules. For subpicomolar concentrations, R exceeds several microns. (b) Time between consecutive single-molecule detection events as a function of the molecular concentration, assuming Brownian diffusion. The markers correspond to experimental data recorded with a 1.2NA microscope objective with a confocal volume of 1.27 fL. The line is a numerical interpolation. (c) Confocal detection volume computed as function of the microscope objective numerical aperture (blue trace, left axis) and concentration where the average number of molecules in the detection volume amounts to 0.01 (red trace, right axis). The confocal detection volume is approcimated to scale with the inverse fourth power of the NA. The reference is taken respective to our 1.2NA microscope objective with a confocal volume of 1.27 fL. (d) Signal to background ratio (SBR) as function of the objective numerical aperture. The reference is taken respective to our 1.2NA microscope objective with a SBR of 100 in the case of Fig. 1c. The signal from a single molecule scales as the excitation intensity times the collection efficiency. Both of these quantities scale as $NA^2$ so that the signal brightness scales as $NA^4$. The background scales as the laser power times the collection efficiency times the total volume, and depends on $NA^{-2}$. We also include a background contribution of 200 counts/s independent of NA to account for the dark counts of the avalanche photodiode.

## S2. Characterization of the influence of the diaphragm diameter

In this section we characterize the influence of the diaphragm diameter (configuration of Fig. 1e) and discuss also how to adequately choose the pinhole diameter. The experiments are based on a FCS calibration using CF640R dye in pure D₂O solution at 1.2 nM concentration, high enough to avoid the need for any background correction or FLCS filtering (Fig. S2a).

The starting point uses the confocal detection volume calibration of 1.27 fL for the 1.2 NA objective when the diaphragm is fully open (Fig. 2a). Using the formula $V_{eff} = \pi^{3/2} w_{XY}^2 w_Z$ determining the confocal detection volume in FCS,[1] we find a lateral waist at 1/e² of $w_{XY} = 357$ nm, which corresponds to a diffusion time of 137 µs for CF640R (we have made the usual approximation that the shape parameter $\kappa = w_Z/w_{XY}$ amounts to 5). From this diffusion time, we calibrate the hydrodynamic radius of CF640R to be 0.92 nm in D₂O. This hydrodynamic radius can then be used to convert the diffusion times $\tau_D = w_{XY}^2/4D$ measured by FCS for various diaphragm diameter (Fig. S2a) into the lateral waist at the laser focus $w_{XY}$ (Fig. S2b). We find that the lateral waist $w_{XY}$ scales as the inverse of the diaphragm diameter $d_{DIAPH}$ as expected from diffraction theory and Fourier optics for diaphragm sizes in the range 0.9 to 4 mm. Due to the Gaussian shape of the laser beam (and the fact that our laser beam diameter slightly under-fills the 6.5 mm back-aperture of the 1.2 NA objective), there is an offset in the dependence of the spot waist with the diaphragm diameter, as $d_{DIAPH}$ values in the range 4 to 6.5 mm yield nearly similar results. The knowledge of the lateral waist $w_{XY}$ can then be expressed as the effective numerical aperture $NA_{EFF}$ for laser excitation using the approximate formula $w_{XY} = 1.2 \lambda / (2 NA_{EFF})$ derived from Abbe's formula. The finding that $w_{XY}$ scales as $1/d_{DIAPH}$ implies that $NA_{EFF}$ is linearly proportional to $d_{DIAPH}$, as expected from a simple geometrical optics consideration (Fig. S2b).

Having determined $w_{XY}$ for each diaphragm size, we can use this knowledge to determine the pinhole diameter. To adequately define the detection volume, the pinhole diameter must accommodate for the laser spot diameter (here 2 $w_{XY}$ times the objective magnification) and an extra size increase (using a prefactor of 2.5 instead of 2x) is often recommended to maximize the fluorescence count rate and avoid an extra loss induced by diffraction is the pinhole size is smaller than the image of the laser spot diameter. Altogether the recommended pinhole diameter is 2.5 $w_{XY}$ times the magnification between the laser focus and the pinhole plane (63x in our case). Due to the limited choice of pinhole diameters commercially available, we have to select the nearest biggest size for our characterization (blue markers in Fig. S2c).

To cross-check the validity of our results, we can compare two independent estimations of the confocal detection volume (Fig. S2d). The first uses the FCS determination of the number of molecules in Fig.

S2a and the known 1.2 nM CF640R concentration (experimental values in Fig. S2d). The second uses the formula $V_{eff} = \pi^{3/2} w_{XY}^2 w_Z$ and the characterization of the lateral spot size $w_{XY}$ in Fig. S2b (thick line in Fig. S2d). Both approaches converge towards similar values, confirming the validity of our calibration. The fact that the point for the smallest 0.9 mm diameter deviates from the $V_{eff}$ prediction can be related to the shape parameter $\kappa = w_Z/w_{XY}$ becoming closer to 8 in this strongly diffracting condition.

Along with the $\tau_D$ and $N$ characterization in Fig. S2a, we also monitor the evolution of the brightness per molecule $CRM$ as a function of the diaphragm diameter, and we record the background intensity $B$ while replacing the CF640R solution with pure D$_2$O (Fig. S2e). Both $CRM$ and $B$ depend on the laser power being used, the corresponding values (measured after the diaphragm) are indicated next to each data point in Fig. S2e. As the diaphragm diameter is reduced, a higher laser power is needed to compensate for the intensity loss due to the increase of the lateral spot size $w_{XY}$. Moreover, the transmission though the diaphragm drops as $d_{DIAPH}$ goes down, requiring even higher laser powers incoming to the diaphragm. Unfortunately, our laser system has a limited power, which in turn leads to a drop of the $CRM$ as the diaphragm diameter is reduced below 3 mm which cannot be totally compensated. Nevertheless, this correspond to the situation used in our main text and can still be used to determine the FLCS figure of merit. As additional comment, we note that the background stems mostly from Raman scattering, and thus its intensity tend to scale with the confocal detection volume. As such, the background intensity strongly increases when the diaphragm diameter is reduced (Fig. S2e).

Lastly, we use the calibration in Fig. S2d,e to compute the FLCS figure of merit $N * CRM/\sqrt{B}$ (Fig. S2f). Here we decide to perform the computation for a concentration of 0.2 pM to ease the comparison with our main text. The blue data (square markers) in Fig. S2f take into account the limited laser power our experimental setup (as indicated on Fig. S2e). The red data (circle markers) indicate the results if the laser power could be increased so that the $CRM$ could be kept constant at the 33,000 counts/s value found for the open diaphragm. These results confirm the finding in the main article that a small diaphragm diameter improves the FoM. Taking into account the power limitations of our setup, the results in Fig. S2f show that the 1.35 mm diameter is a near-optimum case for our system.

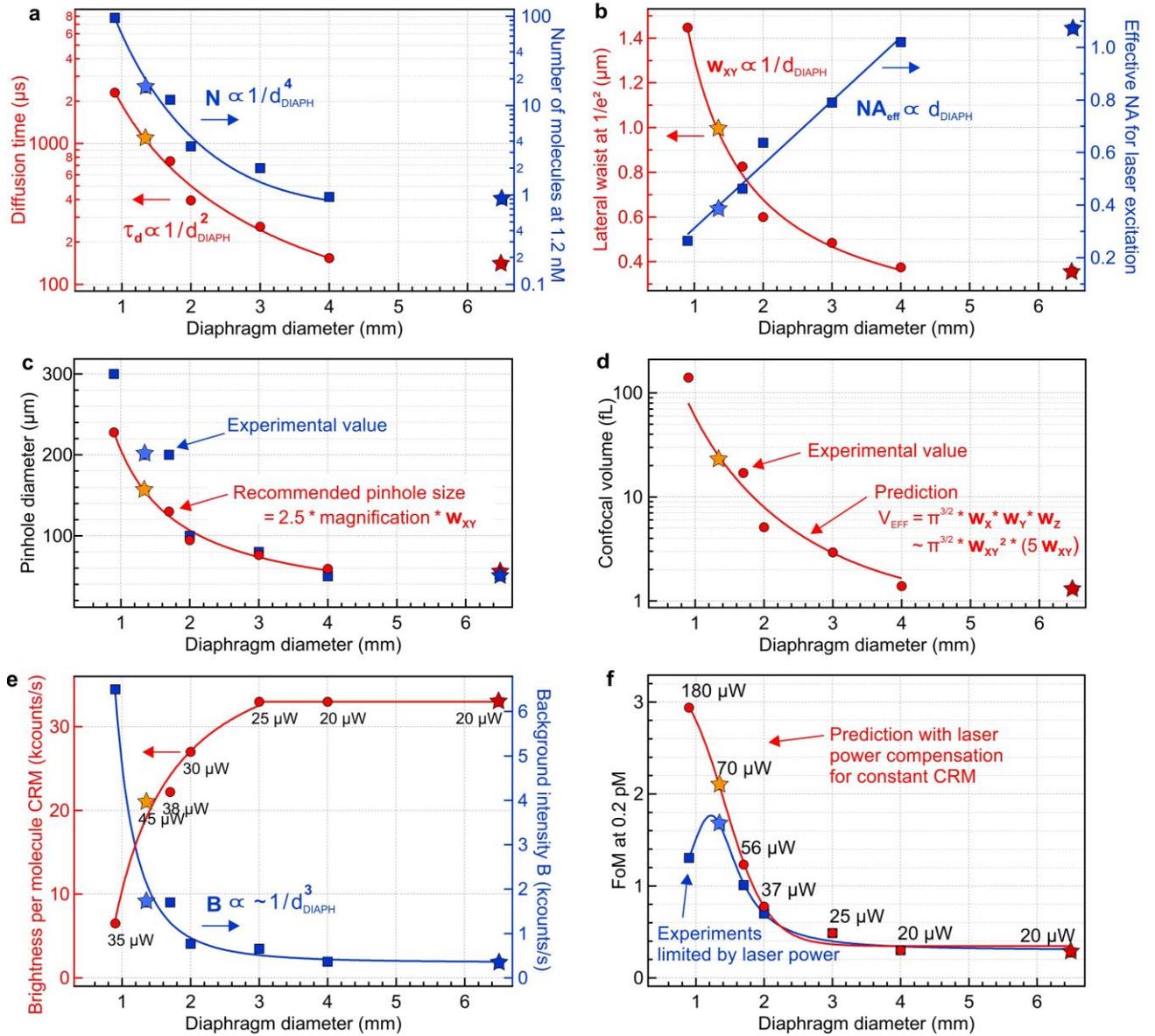

**Figure S2.** (a) FCS characterization of the diffusion time $\tau_D$ and the number $N$ of CF640R molecules as a function of the diaphragm diameter. The CF640R concentration is constant at 1.2 nM. Throughout this figure, the star markers indicate the cases used in the main article: 1.2 NA objective and the 1.2 NA objective with 1.35 mm diaphragm (Fig. 1e). (b) Lateral waist $w_{XY}$ and effective numerical aperture for laser excitation deduced from the diffusion time in (a). (c) Recommended (red circle markers) and used values (blue squares) for the pinhole diameter determined from the lateral waist $w_{XY}$ in (b) and the 63x objective magnification between the laser focus and the pinhole plane. (d) The red markers represent the confocal volume deduced from the FCS number $N$ of CF640R molecules and the concentration in (a). The thick solid line represents the confocal volume deduced from the lateral waist $w_{XY}$ in (b), it is not a fit to the experimental data. (e) Brightness per CF640R molecule $CRM$ and background intensity $B$ as a function of the diaphragm diameter. For each $CRM$ data point we indicate the corresponding 635 nm laser power measured after the diaphragm. (f) FLCS figure of merit $N * CRM/\sqrt{B}$ determined from the data in (a,d,e) for a CF640R concentration of 0.2 pM. Two datasets are presented: for unlimited laser power (red markers, ideal case, the corresponding ideal laser power is indicated next to each marker) and for the constraints of our experimental setup (blue markers, laser powers indicated in (e)).

## S3. Examples of FLCS filters and raw correlation data

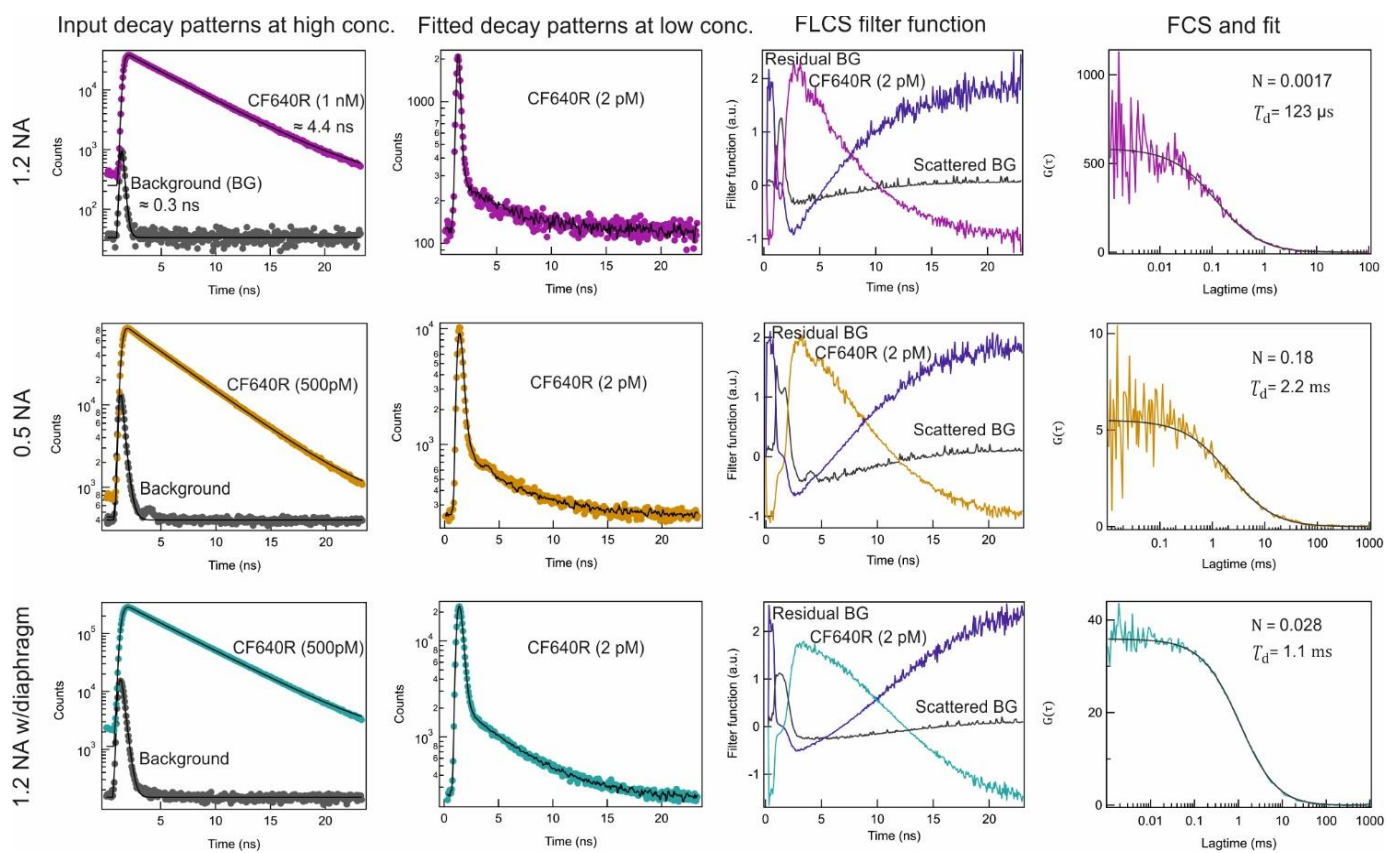

**Figure S3.** Determination of FLCS filters and raw data example at 2 pM concentration. Each line represents a different microscope configuration as illustrated on Fig. 2. The FLCS filters are determined by pattern matching the experimental decays with the reference decays obtained with the CF640R solution at a concentration exceeding 500 pM and the background recorded with the buffer only. This background contains the contribution from scattered excitation photons forming a bump at the beginning of TCSPC decay plus other residual contribution forming the flat line of the TCSPC curve. This ends up into 3 filter functions: CF640R, scattered background and residual background. Only the CF640R filter is used for the signal analysis. The FLCS correlation functions after filtering are shown in the last column for the 2 pM concentration.

## S4. Number of experiments and integration times for the different configurations

| Configuration | Concentration | Number of experiments | Integration time (min) |
|---|---|---|---|
| 1.2 NA with diaphragm | 500-50 pM | 4 | 2 min |
| | 10-1 pM | 5 | 3 min |
| | 0.5 pM | 6 | 4 min |
| | 0.2 pM | 12 | 4 min |
| | 0.1 pM | 6 | 4 min |
| | blank | 12 | 2 min |
| 0.5 NA | 500-50 pM | 5 | 2 min |
| | 10-1 pM | 6 | 3 min |
| | 0.44 pM | 10 | 4 min |
| 1.2 NA | 1000-50 pM | 4 | 2 min |
| | 10-1 pM | 6 | 3 min |
| | 0.5 pM | 7 | 4 min |
| | blank | 12 | 2 min |

**Table S1.** Number of experiments and integration times for the different configurations

## S5. Blank background correlation in absence of target molecule

Here we report experiments performed in the absence of CF640R fluorescent molecules, *i.e.* using only the pure D$_2$O buffer solution. The rest of the experimental conditions are identical to the ones used in Fig. 2a using the 1.2 NA objective with and without the diaphragm. The intensity time trace recorded for the buffer background are processed to compute the FCS or FLCS correlation using the same analysis as for our experiments on CF640R (Fig. S3a,b). Experiments with 2 min integration times are repeated 12 times, allowing us to determine the average correlation amplitude G(0) from the blank and its standard deviation. Due to some potential contamination of unknown origin (fluorescent compounds remaining on the coverslip or the vial, dust or aerosol contaminant…) we still detect some residual correlation even in the absence of the fluorescent target. The concentration of these contaminants appears to be in the 1-10 femtomolar range.

Next, we compare this blank background correlation with the amplitudes recorded in presence of CF640R (Fig. S3c,d). The interpolation of the experimental data with CF640R is based on the following models: for FCS, we use $G(0) = \left(1 - \frac{B}{B+N*CRM}\right)^2 \frac{1}{N}$ which is the classical formula for FCS in presence of background intensity $B$.[1] For FLCS, we extrapolate this formula to $G(0) = \left(1 - \frac{\sqrt{B}/2}{\sqrt{B}/2+N*CRM}\right)^2 \frac{1}{N}$, replacing the background term $B$ in FCS by $\sqrt{B}/2$. While this approach is largely empirical, it provides a good fit for both microscope configurations (with and without the diaphragm) and is based on the rationale of the residual contribution of the shot noise from the background, which scales as $\sqrt{B}$. We use the same approach to fit the relative error in Fig. 2b and obtain a remarkable agreement with the experimental data (see more details in Section S5).

The limit of detection (LOD) can be defined as the dye concentration where the F(L)CS correlation amplitude amounts to the blank background correlation amplitude plus 3 times the standard deviation.[2–4] Likewise, the limit of quantitation (LOQ) corresponds to the blank background plus 9 times the standard deviation.[4] Interpolation of our experimental data with these threshold values indicates a cross-over in the 10 fM range for FLCS for both microscope configurations with and without diaphragm (Fig. S4c,d). This range remains one order of magnitude below the LOQ reported in our main text. Consequently, we can conclude that the false positive and false negative rates remain well below 1% in our experimental data, highlighting the conservative nature of our claims and reinforcing the robustness of our approach.

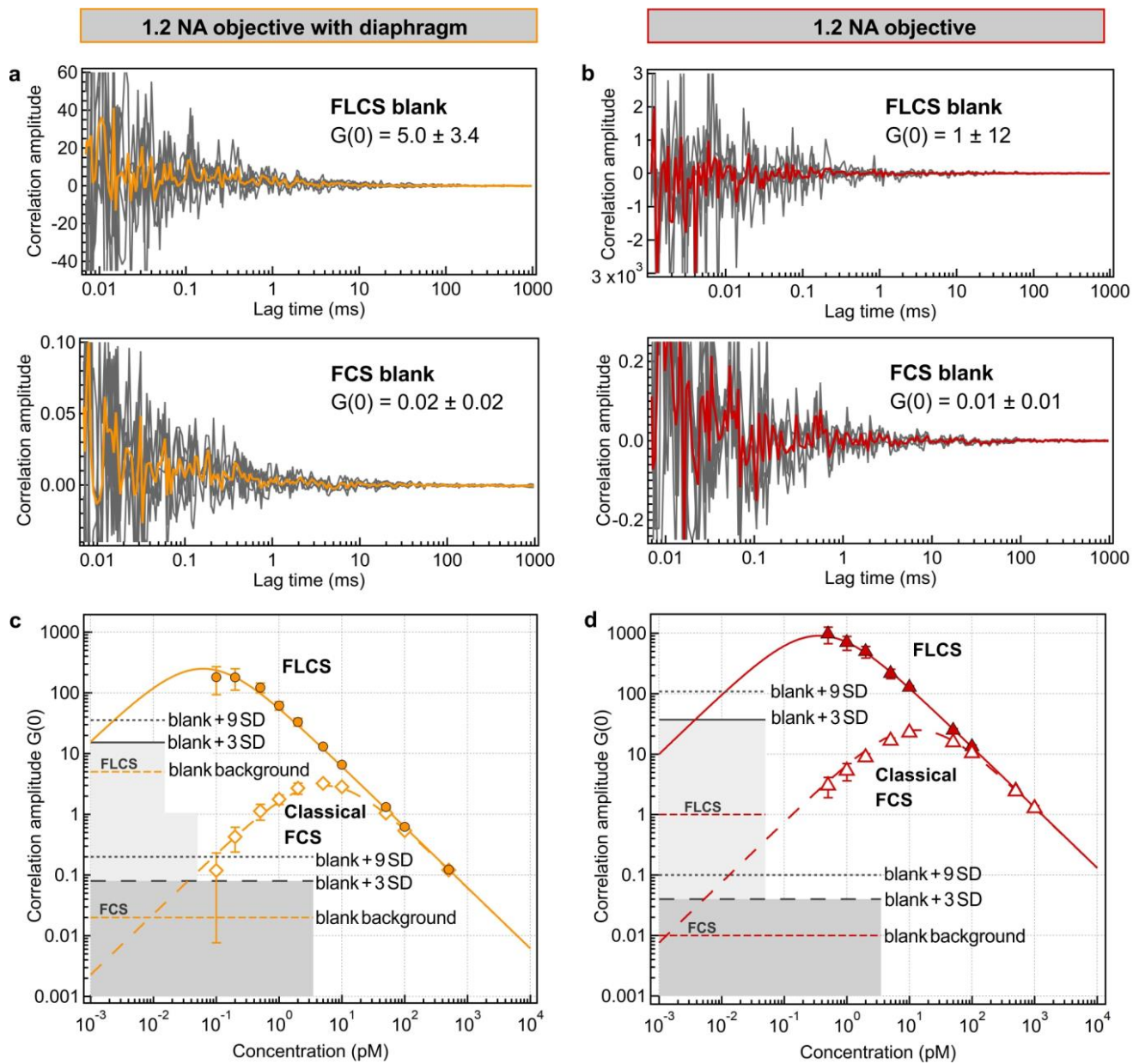

**Figure S4.** (a,b) Blank background FLCS and FCS correlation data recorded for the D$_2$O buffer alone, in absence of CF640R fluorescent molecules. Thin gray curves are individual F(L)CS curves which are fitted using a single species model with free 3D diffusion to determine the correlation amplitude at origin G(0). The thick curves represent the averaged correlation data. The standard deviation is obtained by repeating 12 experiments of 2 min integration time. (c,d) Comparison between the blank correlation amplitudes plus 3 or 9 times the standard deviation (gray horizontal lines, determined in a,b) and the FCS and FLCS correlation amplitudes found in presence of different CF640R concentrations. Markers are experimental data, the error bars represent 2 times the standard deviation. The curves interpolating the data points are computed using the models detailed in the text. The data in a,c refer to the 1.2 NA objective with diaphragm while b,d correspond to the same objective in absence of diaphragm.

## S6. Interpolation of the relative error

We note here $N_{measured}$ the number of molecules measured by FLCS and $N_{expected}$ the ground truth based on the known concentration and confocal detection volume (these can be calibrated at higher concentrations above 10 pM and dilutions are used). From the correction factor introduced in Fig. 3a, the residual shot noise affects the measurement, leading to a change in $N_{measured}$ which can be extrapolated based on the similarity with FCS so that:

$$N_{measured} = \frac{N_{expected}}{\left(1-\frac{\sqrt{B}}{\sqrt{B}+N*CRM}\right)^2} \tag{S1}$$

In this case the relative error $\Delta N = (N_{measured} - N_{expected})/N_{expected}$ can be simplified into:

$$\Delta N = \frac{1}{\left(1-\frac{\sqrt{B}}{\sqrt{B}+N*CRM}\right)^2} - 1 \tag{S2}$$

After some algebra, Eq. (2) can be rewritten into:

$$\Delta N = \frac{\sqrt{B}}{N*CRM}\left(2 + \frac{\sqrt{B}}{N*CRM}\right) \tag{S3}$$

Empirically, we find that a slightly better fit to the data in Fig. 2b can be obtained if we replace the term $\sqrt{B}$ in Eq. (S3) by $\sqrt{B}/2$, adding an extra 50% reduction factor. The origin for this factor remains unknown, but it is already remarkable that such simple approach can achieve a correct interpolation of the data for a range of configurations different by up to two orders of magnitude.

$$\Delta N = \frac{\sqrt{B}/2}{N*CRM}\left(2 + \frac{\sqrt{B}/2}{N*CRM}\right) \tag{S4}$$

Deriving a complete mathematical treatment of residual noise in FLCS is beyond the scope of this paper and the capabilities of its authors. We hope that this work will stimulate more theoretical research into the limits of FLCS.

## S7. The signal to noise ratio in FCS being greater than 1 shows the inaccuracy is a systematic deviation and not a noise problem

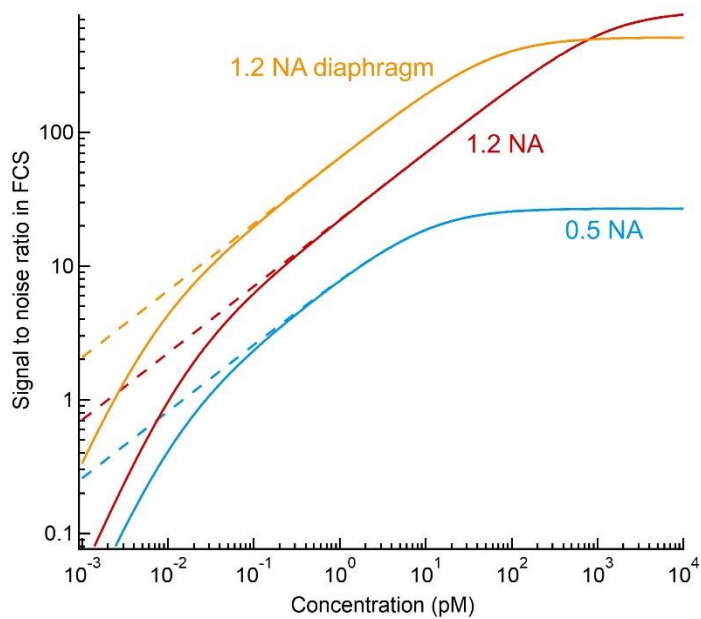

**Figure S5.** Calculation of the signal to noise ratio $SNR$ for the FCS amplitude as a function of the molecular concentration for the different microscope configurations. The $SNR$ in FCS is commonly defined by $SNR = \frac{G(0)}{\sigma(G(0))}$ where $\sigma(G(0))$ is the standard deviation of the correlation amplitude $G(0)$.[5] In presence of background, the FCS signal to noise ratio (solid lines) can be expressed as:[5,6]

$$SNR = CRM \left(1 - \frac{B}{B + N * CRM}\right) \frac{1}{\sqrt{1 + 1/N}} \sqrt{T_{tot}\, \Delta\tau} \tag{S1}$$

where $T_{tot}$ is the total integration time and $\Delta\tau$ the minimum lag time defining the time interval for computing the correlation. For our calculation here we have considered the typical cases with $T_{tot}$ = 60 s and $\Delta\tau$ = 10 μs. The dashed lines show the trend when we set the background $B$ to zero. Note that the expression in Eq. (S1) contains a term $\left(1 - \frac{B}{B+N*CRM}\right) = \left(\frac{SBR}{1+SBR}\right)$ where $SBR = (N * CRM)/B$ is the signal to background ratio, and another term $\frac{1}{\sqrt{1+1/N}}$ to account for the low average number of molecules in the detection volume.[7]

The main result here is that for all the different configurations, we find that the $SNR$ is always greater than 1 for concentrations above 0.05 pM. This implies that the deviations seen experimentally in Fig. 3a,b for concentrations in the range 0.2-2 pM are a systematic error and are not related to some noise problems or statistical imprecision (see also the independence of the error with the integration time Fig. S4).

## S8. The integration time has no influence on the FLCS error

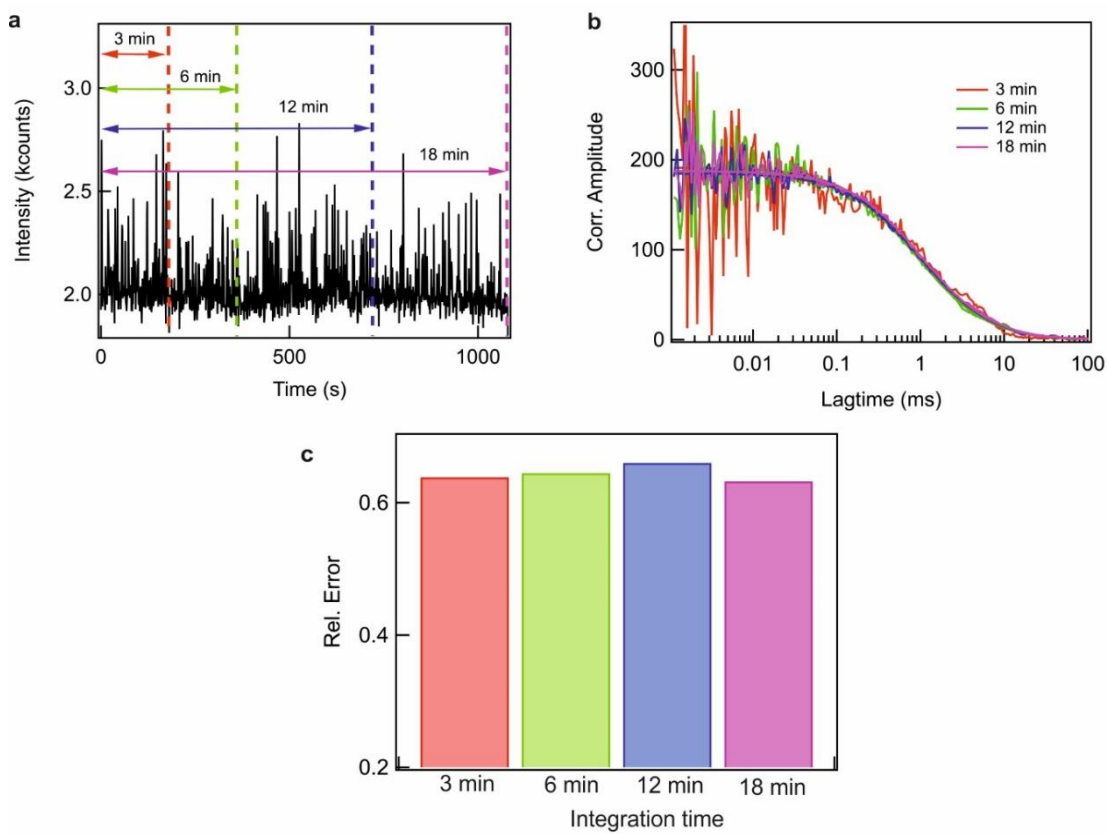

**Figure S6.** (a) Intensity time trace of 0.2 pM CF640R at 45 µW recorded with the 1.2NA objective with diaphragm (Fig. 2c). (b) FLCS correlation data at different integration times $T_{tot}$ as shown in (a). The noise is reduced as $T_{tot}$ increases (as predicted by Eq. (S1)), but the FLCS amplitude remains the same. This means that the relative error in $N$ is unchanged and the accuracy of the FLCS measurement does not depend on $T_{tot}$ in our cases. (c) Histogram showing the relative errors at 0.2 pM for different integration times.

**S9. The lateral flow and the diffusion time have no influence on the FLCS error**

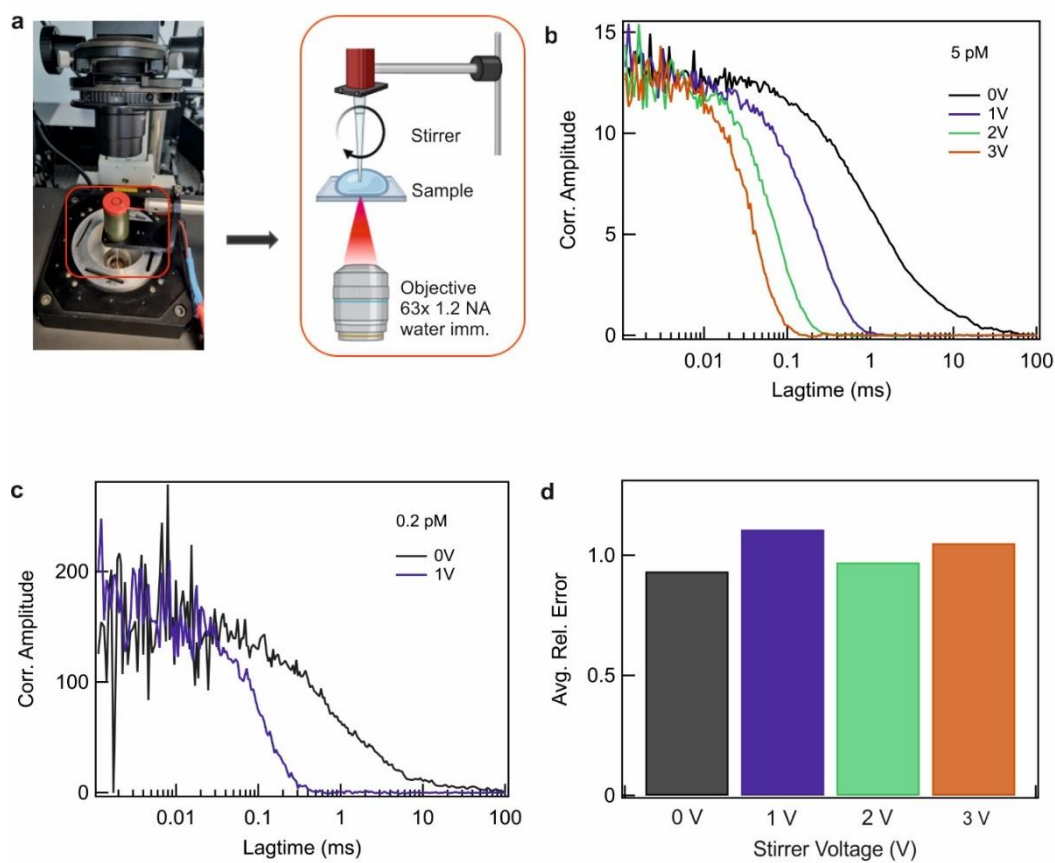

**Figure S7.** (a) Image of the experimental setup used for measurements by mixing the sample using a stirrer and its schematic representation. The speed is regulated by applying different voltages on the rotating motor. The aim is to introduce a lateral flow to speed up the diffusion time and increase the number of events detected per second. The data are taken using the 1.2NA objective with diaphragm (Fig. 2c). (b) Comparison of FCS correlation of CF640R (5 pM concentration) at different stirring speeds (or voltages). We observe a significant decrease in diffusion time (from 1 ms to 35 µs) with increasing stirring speed from 0 V to 3 V. With stirring the CF640R crosses the confocal volume faster which shortens the diffusion time. However, the correlation amplitude remains unchanged as the local concentration (probability of finding a molecule in the detection volume per unit of time) is not modified. (c) Same as (b) for 0.2 pM concentration. Again, the FLCS amplitude remains the same. This means that the relative error in $N$ is unchanged and does not depend on the diffusion time nor on the frequency of detection events in our cases. (d) Histogram showing the relative error at 0.2 pM for different stirring speed.

## S10. TCPSC time gating reduces the FLCS error

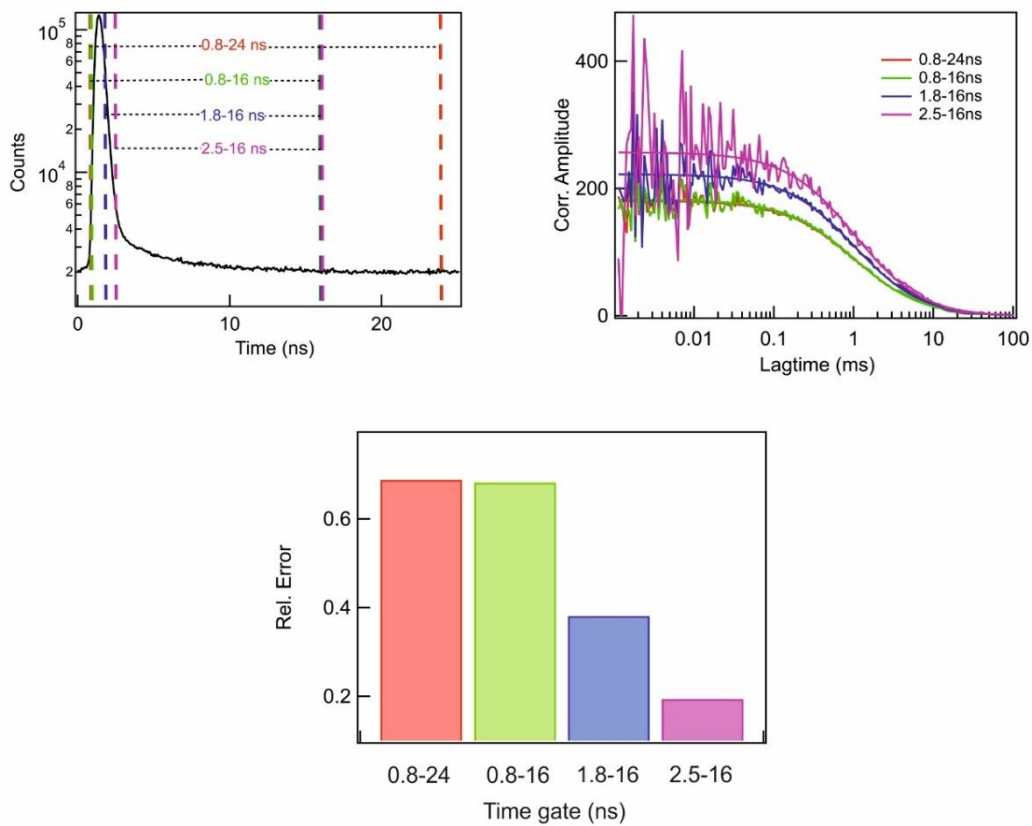

**Figure S8.** (a) TCSPC decay curve of CF640R at 0.2 pM recorded using the 1.2NA objective with diaphragm (Fig. 2c). Different intervals are used to filter and select the photons based on their respective arrival times (time gating).[8] (b) FLCS correlation data at different time gating ranges as shown in (a). The FLCS amplitude increases for narrower time gating intervals, meaning that the relative error in $N$ is affected. (c) Histogram showing the relative errors at 0.2 pM for different ranges of time gating. As time gating is applied, the background level $B$ is reduced and consequently also its residual shot noise fluctuations $\sqrt{B}$ which affects the FLCS accuracy.

## S11. The background noise follows a shot noise dependence Var(B) = B

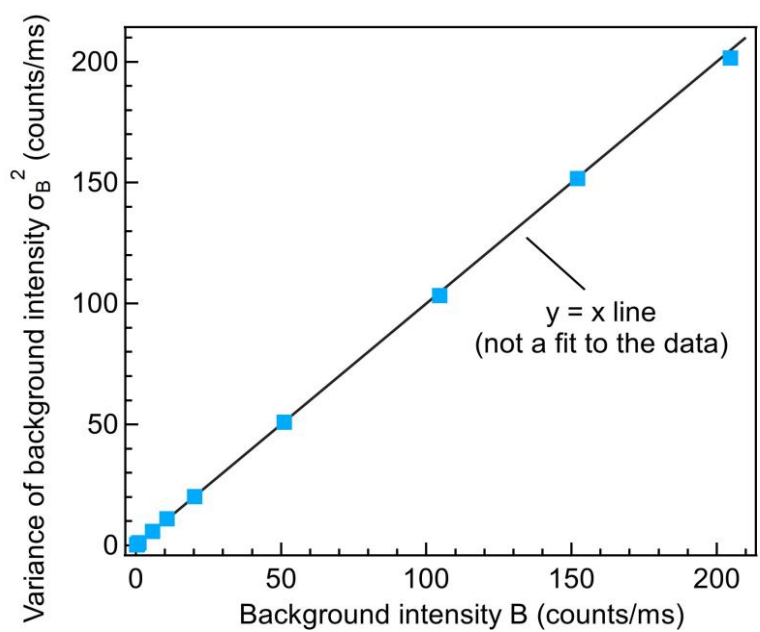

**Figure S9.** Variance (= square of standard deviation) of the background intensity for different background intensities $B$. For this set of experiments, the background level is artificially increased by turning on the microscope halogen lamp (thermal source, no laser nor fluorescence here, 30 s integration time, binning time 1 ms). The linear relationship between $Var(B)$ and the average level $B$ demonstrates that the background noise follows a shot noise dependence determined by a Poisson counting process. As consequence, the standard deviation of the background noise scales as $\sqrt{B}$. This behavior can also be verified for the 3 different microcope configurations of Fig. 1c-e as shown on Tab. S2 below.

| Configuration | Figure | Background $B$ (counts/s) | Standard deviation (counts/s) | Variance $Var(B)$ |
|---|---|---|---|---|
| 1.2 NA objective | 1c | 330 | 19 | 361 |
| 0.5 NA objective | 1d | 1000 | 32 | 1024 |
| 1.2 NA objective with diaphragm | 1e | 1700 | 45 | 2025 |

**Table S2.** Standard deviation of the background level for the different experimental configurations

**S12. The figure of merit is the key factor determining the relative error in the FLCS number of molecules**

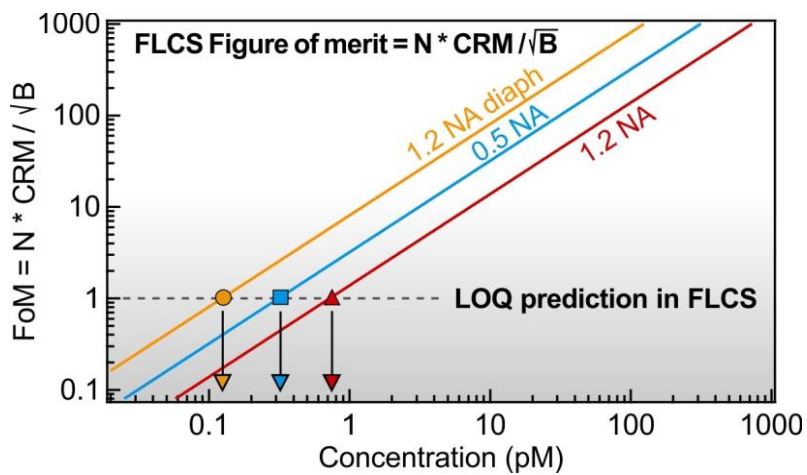

**Figure S10.** FLCS figure of merit $N * CRM/\sqrt{B}$ as function of the concentration for the different microscope configurations. When this FoM becomes lower than 1 (filled markers), the relative error increases strongly and the FLCS technique becomes inaccurate.

## S13. FLCS background correction applied to the 1.2NA microscope configuration

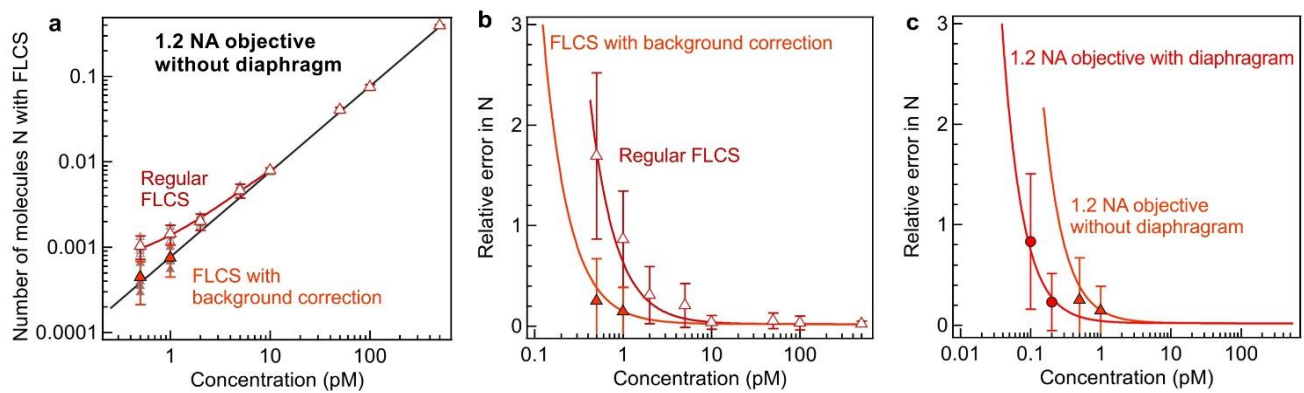

**Figure S11.** FLCS background correction applied to the 1.2NA microscope configuration (without diaphragm, Fig. 2a). (a) Evolution of the average number of molecules measured by FLCS with and without background correction as function of the CF640R concentration. Markers are experimental data, error bars correspond to twice the standard deviation. The black line is the expected trend (ground truth) based on the detection volumes measured by FLCS at concentrations above 10 pM. The color line is a guide to the eye to better visualize the deviation from the linear prediction. (b) Relative error in the number of molecules as function of the concentration with and without background correction determined from the data in (a). Lines are a numerical fit to the data. (c) Comparison between the relative errors after background correction for the 1.2 NA objective with and without diaphragm. The lower error for the diaphragm case (disk markers) shows the configuration in Fig. 2c is superior to the classical confocal case Fig. 2a even after background correction.

## S14. Background correction prefactors for FCS and FLCS

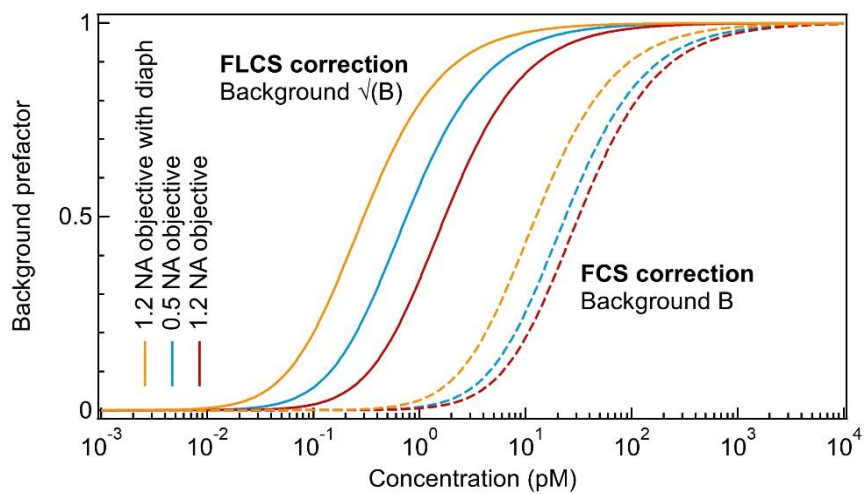

**Figure S12.** Calculation of the FCS and FLCS background correction prefactors to determine the number of molecules as a function of the molecular concentration for the different configurations illustrated in Fig. 2. The classical FCS correction prefactor is $\left(1 - \frac{B}{B+N*CRM}\right)^2$ and by analogy the FLCS correction prefactor introduced in this work is $\left(1 - \frac{\sqrt{B}}{\sqrt{B}+N*CRM}\right)^2$. The background values $B$ are determined experimentally as well as the size of the detection volume. When the correction prefactor is close to 1, the correction is generally not needed and the regular FCS or FLCS are accurate. On the contrary, when the prefactor becomes much smaller, the relative error increases and the uncorrected FCS or FLCS become inaccurate in measuring the number of molecules. These curves are an alternative method to predict when the classical approaches to FCS or FLCS are accurate and the best performing configuration (1.2NA with diaphragm in this case).

## S15. Number of biotin molecules and brightness per molecule as function of streptavidin concentration

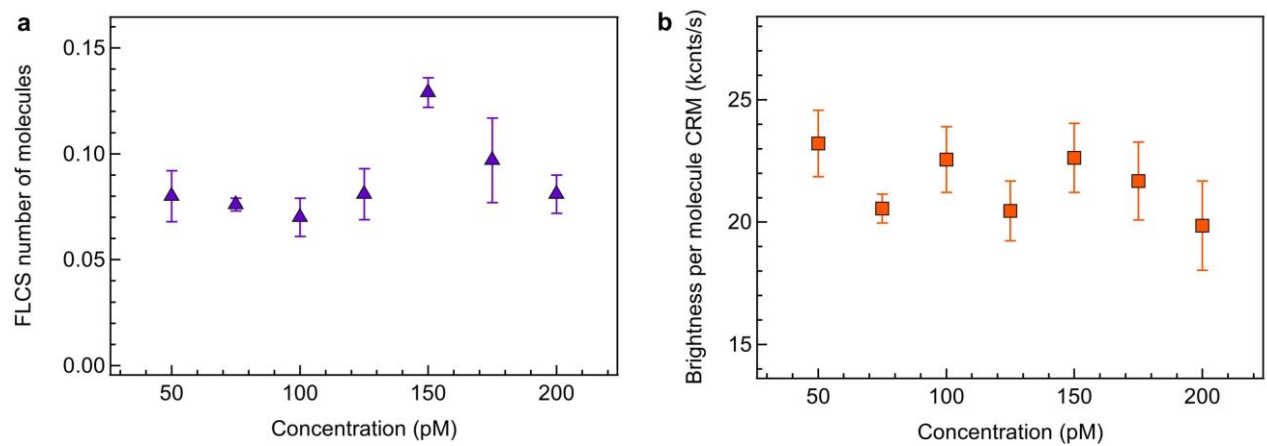

**Figure S13.** (a) FLCS number of biotin molecules and their brightness (b) as a function of the streptavidin concentration for the experiments corresponding to the data in Fig. 5. The relative constant level in both the FLCS number and brightness per molecule serves as a control to ensure the validity of our data.

## S16. Literature values for streptavidin-biotin binding kinetics

| Method | $k_{on}$ (M$^{-1}$·s$^{-1}$) | $k_{off}$ (s$^{-1}$) | Reference |
|---|---|---|---|
| Flow cytometry | (1.3 ± 0.3) × 1e7 | | Buranda et al [9] |
| Droplet microfluidics | 3 e6 to 4.5 e7 | | Srisa-Art et al [10] |
| Silicon nanowires | (5.5 ± 0.08) × 1e8 | | Duan et al [11] |
| Nanophotonics FCS | (3.4 ± 1) × 1e6 | | Tiwari et al [12] |
| Particle diffusometry | (1.74 ± 0.51) × 1e7 | | Ma et al [13] |
| Radiolabelling | | 2.4 e-7 | Piran et al [14] |
| Mass spectrometry | | (4.1 ± 0.1) × 1e-6 | Deng et al [15] |
| Surface plasmon | | (3.3 ± 0.1) × 1e-6 | Pérez-Luna et al [16] |
| FLCS | (2.7 ± 0.2) × 1e7 | << 1e-4 | This work |

**Table S3.** Comparison between the association and dissociation rate constants reported for streptavidin-biotin binding.